\documentstyle[preprint,aps,epsf,epsfig]{revtex}
\tighten

\newcommand{\bq}{\begin{equation}}
\newcommand{\eq}{\end{equation}}
\newcommand{\bqn}{\begin{eqnarray}}
\newcommand{\eqn}{\end{eqnarray}}
\newcommand{\nb}{\nonumber}
\newcommand{\lb}{\label}

\begin{document}

\title{Collapse of a Scalar Field in $2+1$ Gravity}
\author{Eric W. Hirschmann\footnote%
         {Electronic address: \tt ehirsch@kepler.byu.edu\hfil},   
        Anzhong Wang\footnote%
	 {Electronic address: {\tt wang@dft.if.uerj.br}\\  
	    On leave from Departamento de F\'{\i}sica Te\'orica,
            Universidade do Estado do Rio de Janeiro, Brazil\hfil}
}
\address{Department of Physics and Astronomy, Brigham Young University, Provo, Utah 84602}
\author{Yumei Wu\footnote%
         {Electronic address: {\tt yumei@dmm.im.ufrj.br}\\ 
            On leave from Instituto de Matem\'atica,
            Universidade Federal do Rio de Janeiro, Brazil\hfil}
}
\address{Department of Mathematics, Brigham Young University, Provo, Utah 84602}

\date{\today}

\maketitle
 
\begin{abstract}

We consider the problem of critical gravitational collapse of a scalar 
field in $2+1$ dimensions with spherical (circular) symmetry.  After 
surveying all the analytic, continuously self-similar solutions and
considering their global structure, 
we examine their perturbations with the intent of understanding 
which are the critical solutions with a single unstable mode.  The critical
solution which we find is the one which agrees most closely with that 
found in numerical evolutions.  However, the critical exponent which
we find does not seem to agree with the numerical result.  
\end{abstract} 



\newpage

\section{Introduction}

Gravitational collapse is an inherently dynamical and  
multi-dimensional problem.  Its complexity is indirectly
revealed by both the relatively few analytic solutions available for study
and the challenges associated with the construction of a good 
numerical code to simulate the collapse process.  Nonetheless, 
work over the last decade has revealed surprisingly rich behavior in 
even nominally simple ({\it e.g.} spherically symmetric) 
systems\cite{Chop93}.  These 
efforts have led to important advances in our understanding of the 
process of black hole formation and the presence of ``critical" behavior
in these dynamical, gravitating systems.  Indeed, this behavior has been
shown to be present in a great variety of systems\cite{Gun00}.

Some more recent work, from which we take our cue in this paper, has been the  
consideration of the gravitational collapse of a minimally coupled 
scalar field in the presence of a cosmological constant but in a lower 
dimensional spacetime, namely $2+1$.  There are several motivations for 
studying such a  
model beyond the intrinsic interest in examining critical behavior in another
system.  Among these is the recent flurry of work on anti-de Sitter (AdS) 
spacetimes stemming from the AdS/CFT conjecture.  This conjecture posits 
a correspondance between 
the gravitational physics in an AdS spacetime and the physics of
a conformal field theory on the boundary of AdS.  Hence, 
understanding AdS spacetimes can potentially 
yield insight into Super-Yang-Mills theory (and vice versa).  Another 
motivation for studying $2+1$ scalar field collapse is partly the relative 
simplification that results in going from $3+1$ to $2+1$ dimensional 
gravity.  By itself, this wouldn't necessarily be that compelling, but there 
are, of course, some intriguing solutions in $2+1$ such as the BTZ black 
hole that closely parallel the black hole solutions of $3+1$ gravity.  
Earlier work has considered the question of gravitational collapse to a 
BTZ black hole, but using either null fluid or dust as the collapsing 
matter \cite{H94,PS95}.  Thus, considering this particular model of 
gravitational collapse could potentially 
pull together several distinct ideas.

The study of this particular system has included numerical simulations of 
the spherically (or circularly) symmetric model as well as analytic 
investigations.  The numerical results\cite{PC00,HO01} 
have demonstrated the existence of 
critical behavior at the threshold of (BTZ) black hole formation, 
found a continuously self-similar (CSS) critical solution and calculated 
the value of the mass-scaling exponent.\footnote{In \cite{PC00} the exponent 
was found to be $\gamma\approx1.2$ while in \cite{HO01} it was found to be 
$\gamma\approx0.81$.}  On the analytic side, a family of exact solutions to the 
dynamical equations was found under the explicit assumption of 
self-similarity (and zero cosmological constant) \cite{Gar00}.  In addition, 
comparing
these solutions to the numerical results provided strong evidence that one
of the solutions in this family corresponded to the critical solution found
in the simulations.  
This exact solution is interesting, of course, not least because it would appear
to be the first analytic solution which has been shown to correspond with
a critical solution found from numerical simulations.  There is a CSS solution
for the gravitational collapse of a spherically symmetric scalar field in
4D but it is known not to be a critical solution\cite{Rob89}.

Our efforts here are focused on the exact, CSS solutions 
in this model and on identifying 
those that are relevant to critical collapse.  The relative tractability
of $2+1$ gravity allows us greater analytic control of the dynamics in this
system which will hopefully yield to some added insights 
concerning the collapse.  Indeed, as we will see, even the equations
associated with the perturbations of the CSS solutions are integrable.  
As a result, we are able to attempt a fairly exhaustive study of the 
CSS solutions and 
their perturbations.  We begin in the next section with the general
equations of motion for our model.  Section III examines each
of the three classes of CSS solutions and their global structure.  In section 
IV we perturb the exact solutions and examine the stability properties of 
each of these three classes of solutions.  We also calculate the mode 
structure of these solutions and identify the particular exact 
solution which has 
a single unstable mode.  Not surprisingly, this turns out to correspond to 
the same critical solution which Garfinkle found most closely resembled the 
critical solution that appeared in the numerical simulations \cite{Gar00}.
Sections V and VI consider briefly the implications for scalar field collapse 
when there is a potential for the scalar field and offer some conclusions.
An Appendix makes some comments on some more general solutions and properties
of our system in 2+1.
 
In concluding this section, we mention that as our work was 
nearing completion, Garfinkle and Gundlach \cite{GG02} 
presented some recent perturbation results in this same system.  They, too, 
determine analytically the mode structure of these exact solutions with 
an eye to calculating the mass-scaling exponent and confirming that the 
critical solution has a single unstable mode.  As such, there is overlap 
between their work and what we present here.


\section{The Einstein-Scalar Field Equations  }

The action for a scalar field minimally coupled to
$2+1$ gravity is given by\footnote{We will choose units such
that the speed of light is one.} 
\bq
\lb{2.1}   
S = \frac{1}{16\pi G}\int d^3 x \sqrt{-g} \,  \left\{ R - 8\pi G
\left[\nabla_a \phi \nabla^a \phi + 2 V(\phi) +
\frac{\Lambda}{4\pi G}\right]\right\}, 
\eq
where $R$ is the Ricci curvature scalar, $V(\phi)$ is the potential of the 
scalar field $\phi$, and $G$ and $\Lambda$ are respectively the
gravitational and cosmological constants.  The covariant derivative is 
denoted by   $\nabla_{a}$. The gravitational field is then governed by the 
Einstein field equations,  
\bq
\lb{2.2}
R_{ab} - \frac{1}{2}g_{ab}R + \Lambda g_{ab} = 8\pi G T_{ab},
\eq
with the energy-momentum tensor given by
\bq
\lb{2.3}
T_{ab} = \nabla_{a}\phi \nabla_{b}\phi  -
\frac{1}{2}g_{ab}\left[\nabla_{c}\phi \nabla^{c}\phi +  2 V(\phi)\right].
\eq
The evolution of the scalar field is determined by 
\bq
\lb{2.5}
\nabla_a\nabla^a \phi = \frac{\partial V(\phi)}{\partial \phi}.
\eq

\begin{figure}[htbp]
\begin{center}
\label{fig1}
\leavevmode
    \epsfig{file=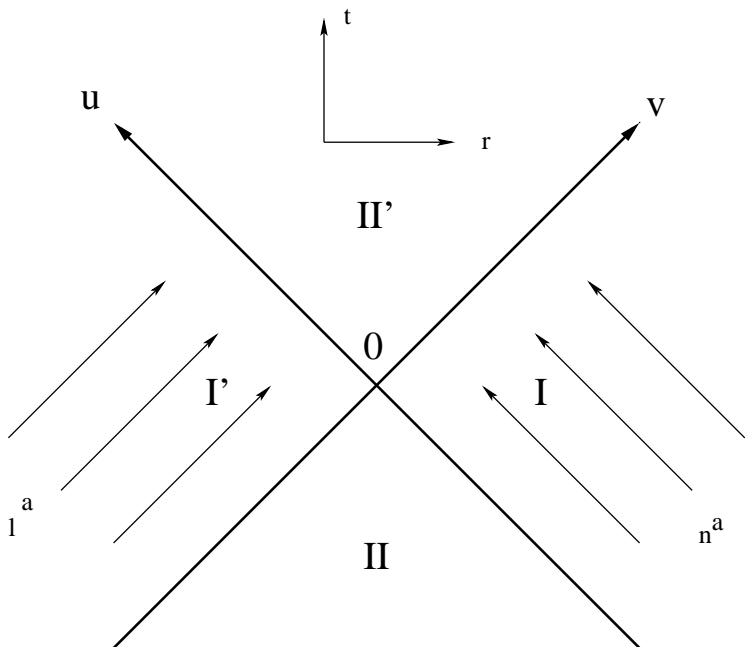,width=0.6\textwidth,angle=0}
\caption{The spacetime in the ($u, v$)-plane. Each of the four regions
is defined as follows:  $I \equiv \{x^{a}: u \le 0,\; v \ge 0\}$, 
$II \equiv \{x^{a}: u
\le 0,\; v \le 0\}$, $ I' \equiv \{x^{a}: u \ge 0,\; v \le 0\}$
and $II' \equiv \{x^{a}: u \ge 0,\; v \ge 0\}$.  The null geodesics, 
$l^a$, lying along the hypersurfaces $u=$ const. are outgoing, while 
the null geodesics, $n^a$, lie along the hypersurfaces $v=$ const. and are 
ingoing.}
\end{center}  
\end{figure}

The general metric for a ($2+1$)-dimensional spacetime with spherical
symmetry\footnote{Sometimes the spacetimes
described by metric (\ref{2.6}) are also referred to as having axial
or circular symmetry.} can be cast in the form,
\bq
\lb{2.6}
ds^2 = - 2 e^{2\sigma(u,v)} du dv + r^2(u,v) d\theta^2,
\eq
where  $(u,v)$ is a pair of null coordinates varying in the range
$(-\infty,\infty)$, and $\theta$ is the usual angular coordinate with the
hypersurfaces $\theta = 0, \; 2\pi$ being identified.  It should be noted
that the form of the metric is unchanged under the coordinate transformations,
\bq
\lb{2.6aa}
u = u(\bar{u}),\;\;\;\; v = v(\bar{v}).
\eq
In the following we will use this gauge freedom to fix some of the integration
constants which arise.  In addition, the roles of $u$ and $v$ can be
interchanged. To fix this particular freedom, we choose coordinates such
that along the lines of constant $u$
the radial coordinate $r$ increases towards the future, while along the
lines of constant $v$ the coordinate $r$ decreases towards the
future.  This, of course, just defines $u$ as outgoing and $v$ as ingoing null
coordinates [{\it cf}. Fig.1].

Defining  null vectors, $l_{a}$ and $n_{a}$, along each of the two rays by   
\bq 
\lb{2.6a}
l_{a} = \delta^{u}_{a},\;\;\;\;
n_{a} = \delta^{v}_{a},
\eq
we find that  
\bq 
l^{b}\nabla_{b}l_{a}   = 0 = n^{b}\nabla_{b}n_{a},    
\eq
{\it i.e.} these  null rays  are  affinely parameterized null geodesics. 
The expansion for each is given by  
\bqn
\lb{2.6b}
\theta_{l} &\equiv& - \nabla_{\lambda}l^{\lambda}  =
e^{-2\sigma}\frac{r_{,v}}{r},\nb\\
\theta_{n} &\equiv& -  \nabla_{\lambda}n^{\lambda} =
e^{-2\sigma}\frac{r_{,u}}{r},
\eqn
where, as usual, the comma notation denotes partial differentiation.  

The corresponding Einstein-scalar
field equations (\ref{2.2}) for the metric (\ref{2.6}) take the form,
\bqn 
r_{,uu} - 2\sigma_{,u} r_{,u} &=& - 8\pi G r \phi_{,u}^{2},\\ 
r_{,vv} - 2\sigma_{,v} r_{,v} &=& - 8\pi G r \phi_{,v}^{2},\\ 
r_{,uv} + 2r\sigma_{,uv}  &=& - 8\pi G r \left(\phi_{,u}\phi_{,v} 
- 2e^{2\sigma}\tilde{V}\right),\\
\lb{2.7}
r_{,uv}    &=&  8\pi G r  e^{2\sigma}\tilde{V},
\eqn
where   
\bq
\lb{2.4}
\tilde{V}(\phi) \equiv V(\phi) + \frac{\Lambda}{8\pi G}.
\eq 
The equation of motion for the scalar field (\ref{2.5}) is now
\bq
\lb{2.8}
 2\phi_{,uv} + \frac{1}{r}\left(r_{,u}\phi_{,v} + r_{,v}\phi_{,u}\right)
= -  e^{2\sigma}\frac{\partial V(\phi)}{\partial \phi}.
\eq
Because of the presence of the effective potential $\tilde{V}(\phi)$,   the 
above equations do not, in general, allow the existence of exactly
self-similar solutions. However, because our main interest is in  
phenomena occuring  at the threshold of black hole formation,
we may still seek for solutions that are asymptotically self-similar 
\cite{GMG96}. Indeed,  what we
are really looking for in the study of (Type II) critical collapse is a   
solution that approaches  self-similarity as the collapse is about to
form a spacetime singularity, that is, in the limit of small spacetime scale.
To see the above clearly, let us first introduce the
dimensionless variables, $x$ and $\tau$, via the relations 
\bq
\lb{2.9}
x  = \ln\left(-\frac{v}{u}\right),\;\;\;\;
\tau = - \ln\left(-\frac{u}{u_{0}}\right),
\eq
where $u_{0}$ is a dimensionful constant, and the above relations are assumed to
be valid only in the region $v\ge 0,\; u \le 0$.  We will refer to this region 
as
Region I ({\it cf}. Fig 1.).  Later, we will consider extensions of these
coordinates to other
regions of the ($u,v$)-plane. In terms of $x$ and $\tau$, the Einstein 
field equations (\ref{2.7}--\ref{2.8}) take the form  
\bqn
\lb{2.13a}
s_{,x x } + 2s_{,\tau x } + s_{,\tau\tau} - s_{,x } -
s_{,\tau} &=&   2(\sigma_{,x } + \sigma_{,\tau})(s_{,x } +
s_{,\tau} - s) 
\nb\\
          & & \quad - 8\pi G s \left(\varphi_{,x } + \varphi_{,\tau} -
c\right)^{2} \\
\lb{2.13b}
s_{,x x } - (1 + 2 \sigma_{,x }) s_{,x }  &=& - 8\pi G s
\varphi_{,x }^{2} \\ 
\lb{2.13c}
s_{,x x } + s_{,\tau x }   - s_{,x } 
+ 2s\left(\sigma_{,x x } + \sigma_{,\tau x }\right)  
   &=& - 8\pi G s \Big[  \varphi_{,x }\left(    \varphi_{,x } 
                                               + \varphi_{,\tau} 
                                               - c
                                      \right) 
\nb\\
& & \qquad\qquad
                        - 2u^{2}_{0} \, e^{-2\tau + 2\sigma + x } 
                                     \, \tilde{V}(\phi)
                \Big] \\
\lb{2.13d}
s_{,x x } + s_{,\tau x }   - s_{,x }    
   &=&  8\pi G s u^{2}_{0} \, e^{-2\tau + 2\sigma + x } \, \tilde{V}(\phi) \\
\lb{2.13e}
2s\left(\varphi_{,x x } + \varphi_{,\tau x }\right) +
\varphi_{,x }\left(2s_{,x } + s_{,\tau} - s\right) 
   &=&  - \varphi_{,\tau}s_{,x } + c s_{,x }    \nb\\
   & & \quad - s u^{2}_{0} \, e^{-2\tau + 2\sigma + x } 
                     \, \frac{\partial {V}(\phi)}{\partial \phi}, 
\eqn
where
\bqn
\lb{2.14}
r(u,v) &\equiv& - u s(u,v),\nb\\
\phi(u,v) &\equiv&  c \ln\left|-u\right| + \varphi(u,v),
\eqn
with $c$ being an arbitrary constant.

In every example of Type II critical collapse studied so far, the critical solutions have 
been found to have either discrete self-similarity (DSS) or continuous 
self-similarity (CSS) \cite{Gun00}.  Discrete self-similarity of the 
spacetime corresponds to periodicity of solutions in the coordinate $\tau$, 
 \bq
\lb{3.1}
{\bf A}(\tau, x ) = {\bf A}(\triangle + \tau, x ),
\eq
where the vector ${\bf A}$ is defined as ${\bf A} = \{\sigma, \; s,\; 
\varphi\}$,
and the constant $\triangle$ denotes the period of the solutions, while
continuous self-similarity corresponds to the case in which  
\bq 
\lb{3.2}
{\bf A}(\tau, x ) = {\bf A}(x ),
\eq
with the homothetic Killing vector $\xi^{a}$ being given by
\bq 
\lb{3.3}
\xi ^{a} = u\delta^{a}_{u} + v \delta^{a}_{v}.
\eq

In our current problem, we can see from Eqs.~(\ref{2.13a}--\ref{2.13e}) 
that the field equations contain an explicit factor of $e^{\tau}$ and
consequently neither of the two kinds, DSS and CSS, of self-similar solutions 
is allowed. The physical reason, of course, is that the presence of the 
effective potential
introduces a scale into the problem and, as a result, the field
equations are not scale invariant.  However, as mentioned previously,  
critical collapse is only relevant in the region where $\tau \gg 1$
($0 < -u \ll 1$).  Therefore, provided that 
\bq
\lb{3.4}
e^{-2\tau} \tilde{V}(\phi) \rightarrow 0,
\eq
as $\tau \rightarrow \infty$, solutions that asymptotically approach 
self-similar ones will exist, and will be well approximated by
the self-similar solutions that satisfy Eqs.(\ref{2.13a}--\ref{2.13e}) with 
$\tilde{V}(\phi) = 0$. This condition holds for several physically 
interesting cases.  One such case is 
a massless scalar field in an AdS spacetime background in
which we have $\tilde{V} = \Lambda/8\pi G$.  Another is the potential for weakly
interacting pseudo-Goldston bosons \cite{HSF89}, for which we have  
$$
V(\phi) = V_{0}\cos^{2(1-p)}\left(\frac{\phi}{f(p)}\right),\;\;\;\;
f(p) = \left[\frac{p(1-p)}{4\pi G}\right]^{1/2},
$$
where $V_{0}$ and $p$ are constant, and $0 < p < 1$.

Indeed, as mentioned in the introduction, recent numerical 
simulations \cite{PC00,HO01} provide strong evidence that 
the critical solution in the case $\tilde{V} = \Lambda/8\pi G$ is 
continuously self-similar.  This suggests the curious result that although 
the cosmological 
constant is crucial for the formation of a BTZ black hole, the actual
critical solution does not depend on it. 
This, then, will serve as our motivation to only consider CSS solutions 
in the following sections.

\section{Solutions With Continuous Self-Similarity}

In order to construct CSS solutions, we must do more than assume that Eq.(\ref{3.4})
holds in the limit as $\tau\rightarrow\infty$.  We will assume it as an
identity and therefore set $\tilde{V} = 0$ in Eqs.(\ref{2.13a}--\ref{2.13e}).  Doing
this, we can drop all dependance on $\tau$.  We also change variables
by choosing to use the somewhat more common similarity variable, $z=e^x$.  
Making these substitutions in the field equations, we get 
\bqn 
\lb{3.5a}
 z^{2}s'' + 2z\sigma'(s - zs') &=& -8\pi G s(c -
z\varphi')^{2},\\ 
\lb{3.5b} 
s'' - 2\sigma's' &=& - 8\pi G s {\varphi'}^{2}, \\
\lb{3.18c}
zs'' + 2s(z\sigma'' + \sigma') &=& 8\pi G s\varphi'(c - z\varphi'),\\ 
\lb{3.5d}
s'' &=& 0,\\
\lb{3.5e}
2zs\varphi'' + \varphi'(2zs' + s) - c s' &=& 0,
\eqn
where a prime denotes ordinary differentiation with respect to $z$. From 
Eq.(\ref{3.5d}) we
find that 
\bq
\lb{3.7}
s(z) = a_{0}z - b_{0},
\eq
where $a_{0}$ and $b_{0}$ are integration constants.  Depending on the
values of these two constants, the solutions can have very different physical
interpretations. Thus, we will consider them separately in the following.

\subsection{$a_{0} \not= 0$ and $ b_{0} = 0$}

For the case in which $s(z) = a_0 z$, we find that
Eqs.(\ref{3.5a}--\ref{3.5e}) have the general solution,  
\bq
\lb{3.8}
\sigma(z) =  \frac{\chi}{2}  \ln\left|z\right| + \sigma^{1}_{0},\;\;\;
\varphi(z) =  c \ln\left|z\right| +\varphi^{1}_{0},\;\;\; (b_{0} = 0), 
\eq
where $\sigma^{1}_{0}$ and $\varphi^{1}_{0}$ are integration constants with 
$\chi \equiv 8\pi c^{2}G \ge 0$. By rescaling $u$ and $v$ and redefining the 
constant $\varphi^{1}_{0}$,   we
can always set $a_{0} = 1$ and $\sigma^{1}_{0} = 0$, a condition that we
shall assume in the following discussion. 
Introducing the two new coordinates $\bar{u}$ and $\bar{v}$ 
via the relations  
\bq
\lb{3.10} 
d\bar{u} = \frac{du }{(1+\chi)(- u)^{\chi}} ,\;\;\;\;\;
d\bar{v} = (1+\chi){v}^{\chi} d{v},
\eq
we find that the corresponding metric and the massless scalar field are given,
respectively, by
\bqn
\lb{3.11a}
ds^{2} &=& - 2d\bar{u}d\bar{v} + \bar{v}^{2/(1+\chi)}d\theta^{2},\nb\\
\phi &=& \frac{c}{1+\chi} \ln\left(\bar{v}\right) + \phi^{1}_{0}, 
\eqn
where $\phi^{1}_{0}$ is another constant. 
From Eq.(\ref{3.11a}) we find that
\bq
\lb{3.13}
\nabla_{a}\phi = \frac{c}{(1+\chi)\bar{v}}\delta^{\bar{v}}_{a},\;\;\;\;
R_{ab} = 8\pi G  \left(\frac{
c}{(1+\chi) \bar{v}}\right)^{2}\delta^{\bar{v}}_{a}\delta^{\bar{v}}_{b},
\eq
which shows that the massless scalar field is equivalent to 
an ingoing null dust flowing along the null hypersurfaces $\bar{v} =$ const.  
Because the hypersurface $\bar{u} = 0$ is regular, to have a
geodesically maximal spacetime, we must extend the spacetime beyond this
surface. One simple analytic extension is to take the range of $\bar{u}$ in
Eq.(\ref{3.11a}) simply as $\bar{u} \in (-\infty,\, +\infty)$.  On the other 
hand, the hypersurface $\bar{v} = 0$ is singular, and, as shown in 
Appendix A, the nature of the singularity is strong
in the sense that the distortion experienced by a freely falling observer
becomes infinitely large at $\bar{v} = 0$ ($v = 0$).  The corresponding 
Penrose diagram is given in Fig. 2.

\begin{figure}[htbp]
\begin{center}
\label{fig2}
\leavevmode
    \epsfig{file=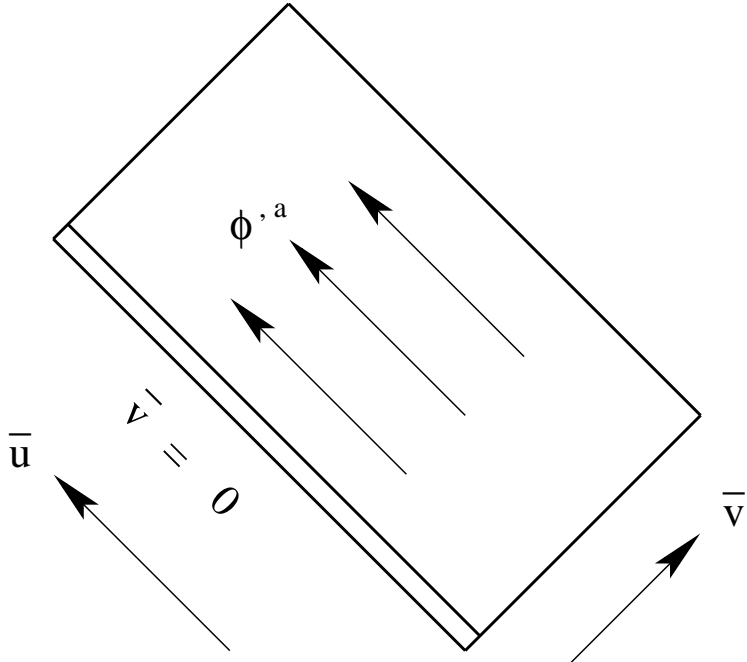,width=0.6\textwidth,angle=0}
\caption{The Penrose diagram for the solutions described by Eq.(\ref{3.11a}) 
in the text, where the double line at $\bar{v} = 0$ ({\i.e.} $\; r = 0$) 
represents the spacetime singularity.} 
 \end{center}   
\end{figure}  

\subsection{$a_{0} = 0$ and $ b_{0} \not= 0$}

In the case where $s(z) = - b_0$, we find that
Eqs.(\ref{3.5a}--\ref{3.5e}) have the {general solution},  
\bq
\lb{3.16}
\sigma(z) =  - \frac{\chi}{2}  \ln\left|z\right| + \sigma^{1}_{0},\;\;\;
\varphi(z) =   \varphi^{1}_{0},\;\;\; (a_{0} = 0). 
\eq
Again, by a rescaling of $u$ and $v$, we can set $b_{0} = -1$ 
and $\sigma^{1}_{0} = 0$. 
In a manner completely analagous to the previous case, we introduce two new 
coordinates $\bar{u}$ and $\bar{v}$, but defined in the current case by
\bq
\lb{3.17} 
d\bar{u} = (1 + \chi)(-u)^{\chi} du, \;\;\;\;\;
d\bar{v} = \frac{d{v}}{(1 + \chi) {v}^{\chi}}.
\eq
Doing so, we find that the corresponding metric and scalar 
field are given, respectively, by   
\bqn
\lb{3.18a}
ds^{2} &=& - 2d\bar{u}d\bar{v} + \bar{u}^{2/(1 + \chi)}d\theta^{2},\nb\\
\phi &=& \frac{c}{1 + \chi} \ln\left(- \bar{u}\right) + \phi^{1}_{0}.
\eqn
Note that the above metric can be obtained directly from Eq.(\ref{3.11a})  
by exchanging the two null coordinates $\bar{u}$ and $\bar{v}$.    
Thus, these two spacetimes must have the same local and global
properties after such an exchange takes place.  In particular,
the massless scalar field is again energetically equivalent to that of
null dust, but this time flowing outwards along the null hypersurfaces 
defined by $\bar{u} = $ const.  Moreover, the hypersurface $\bar{u} = 0$
is indeed singular for all values of $c$ and, as shown in Appendix A, 
and in contrast to the claims made in \cite{CF01}, the
singularity is strong in the sense that the distortion experienced by a 
freely falling observer becomes unbounded as this surface is approached. The 
corresponding Penrose diagram is given in Fig. 3. 

\begin{figure}[htbp]
\begin{center}
\label{fig3}
\leavevmode
    \epsfig{file=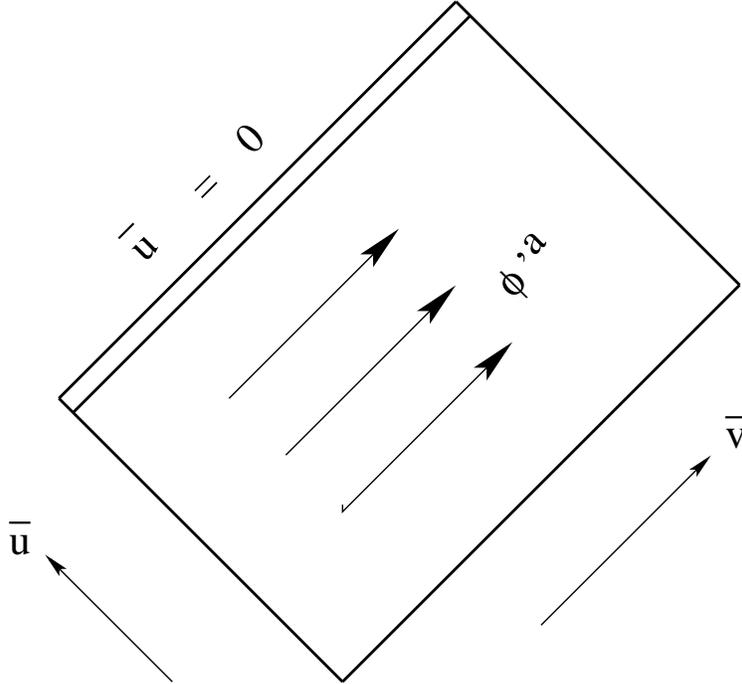,width=0.6\textwidth,angle=0}
\caption{The Penrose diagram for the solutions described by 
Eq.(\ref{3.18a}) in the text, where the double line at $\bar{u} = 0$ 
represents the spacetime singularity. } 
 \end{center}   
\end{figure}  

We also note that in \cite{CF01} the validity of the solutions of 
Eq.(\ref{3.18a}) was restricted to the region $\bar{u} \le 0,\; \bar{v} 
\ge 0$. However, in order to obtain a geodesically maximal spacetime, 
as we have done here, 
it is necessary  to extend these solutions to the region $\bar{v} \le 0$. This 
observation will be crucial when we consider the boundary conditions for 
perturbations in section IV.  

\subsection{$a_{0}b_{0} \not= 0$}

When $a_{0}b_{0} \not= 0$, it can be shown from Eqs.(\ref{3.5a}--\ref{3.5e})
that the constant $b_{0}/a_{0}$ must be non-negative,
and that the general solution is given by   
\bqn 
\lb{3.22}
\sigma(z) &=&  2\chi
\ln\left|{z}^{1/4} + \alpha {z}^{-1/4}\right| + \sigma^{1}_{0} \nb\\
\varphi(z) &=&  2 c 
\ln\left|{z}^{1/2} + \alpha \right| + \varphi^{1}_{0},  \;\;\;\; 
(a_{0}b_{0} \not= 0) 
\eqn
where  
\bq
\lb{3.23}
\alpha \equiv \pm \sqrt{{b_{0} \over a_{0}}}
\eq
and, as before, $\chi = 8\pi G \, c^2$.  For 
$\alpha = 0\;$ or $\alpha =\pm \infty$, the solutions reduce to those
studied in the last two subsections.  Thus, in the rest of this subsection 
we will only consider solutions with finite, nonzero $\alpha$.  
Again, if we rescale the two null coordinates and redefine the 
constant $\alpha$, we can always set  
\bq
\lb{3.23a}
a_{0} = - 1,\;\;\;\;
\sigma^{1}_{0} = -\frac{1}{2} \ln\left(2^{4\chi} \alpha^{2\chi}\right) 
\eq
so that at the origin of coordinates, the coordinate $u$ defines proper time.   

In the case that $\alpha = 1$, the corresponding solutions reduce to those
originally found by Garfinkle \cite{Gar00}.  In addition, he found that 
in the strong field regime the particular solution with   
$\chi_{crit} = 7/8$ is very similar to the critical solution found 
numerically by Pretorius and Choptuik \cite{PC00}. 
When $\alpha = -1$ the corresponding solutions reduce to those
given in \cite{CF01}.

In terms of $u$ and $v$, the metric coefficients and the massless scalar field
are given by
\bqn
\lb{3.25}
e^{2\sigma} &=&  \frac{1}{(16\alpha^2)^\chi}\frac{\left[\alpha
(-u)^{1/2} + v^{1/2}\right]^{4\chi}}{(- u v)^{\chi}} \nb\\
r(u,v) &=& \alpha^{2}(-u) - v \nb\\
\phi(u,v)&=& 2c\ln\left|\alpha(-u)^{1/2} + v^{1/2}\right| + \phi^{1}_{0} 
\eqn
from which we find that
\bq
\lb{3.26}
\nabla_{a}r\nabla^{a}r = - 2\alpha^{2}e^{-2\sigma} \le 0,
\eq
{\it i.e.} $\nabla_{a}r$ is non-spacelike.  The curvature is readily found 
from Eq.(\ref{3.25}) to be  
\bq
\lb{3.27}
R \equiv R_{ab}g^{ab} = 8\pi G \nabla_{a}\phi\nabla^{a}\phi  
= 2 \alpha \chi (16\alpha^2)^\chi \, \frac{(-uv)^{\chi - 1/2}}
{\left|\alpha (-u)^{1/2} + v^{1/2}\right|^{2(2\chi + 1)}}.
\eq
 From this, we can see that, for $\chi \ge 1/2$, although the metric 
coefficients are singular along both of the hypersurfaces 
$v = 0$ and $u = 0$, the curvature is perfectly regular there.
We must therefore 
extend the solutions beyond these surfaces.  

Note, however that in the case that $0 < \chi < 1/2$, these 
hypersurfaces at $v = 0$ and $u = 0$ are singular, and their physical 
interpretation becomes unclear.  For that reason, we will only consider
the case $\chi \ge 1/2$ in the following.  In order to be somewhat 
systematic in our further study of these solutions, we will consider 
each of the three subcases $ \chi = 1/2,\; 1/2 < \chi < 1$ and $\chi \ge 1$, 
separately.  Finally, it is interesting to note that for all values 
of $\chi$, the spacetime is singular at the point $(u, v ) = (0, 0)$.  

\subsubsection{ $ \chi = \frac{1}{2}$}

For $\chi = 1/2$, we introduce two new coordinates, $\bar{u}$ and $\bar{v}$, 
via the relations,
\bq
\lb{3.28}
\bar{u} = - \sqrt{-u} ,\;\;\;\;\;
\bar{v} =  \sqrt{v},
\eq
and find that in terms of $\bar{u}$ and $\bar{v}$ the metric and the
massless scalar field are given by
\bqn
\lb{3.29}
ds^{2} &=& - \frac{2}{|\alpha|} \, f^{2}(\bar{u},\bar{v})d\bar{u}d\bar{v} +
r^{2}(\bar{u},\bar{v})d\theta^{2} \nb\\ 
r &=& \alpha^{2}(-\bar{u})^{2} - \bar{v}^{2}  \nb\\
\phi &=& 2c\ln\left|f(\bar{u},\bar{v})\right| + \phi^{1}_{0} 
\eqn
where  
\bq
\lb{3.30}
f(\bar{u},\bar{v}) \equiv   \alpha (-\bar{u}) + \bar{v}.
\eq
It is straightforward to show 
\bq
\lb{3.31}
\nabla_{a}r\nabla^{a}r  = 8\alpha^{2} \, \bar{u}\bar{v} \, e^{-2\sigma} 
\eq
namely, that in Region I where $\bar{u} \le 0$ and $\bar{v} \ge 0$, the
hypersurfaces $r(\bar{u}, \bar{v}) = $ const. are always spacelike,   
including the one at $r(\bar{u}, \bar{v}) = 0$ which forms the upper  
boundary of the spacetime [{\it cf}.  Fig. 4]. 

Using Eq.(\ref{3.29}) to calculate the ingoing and outgoing expansions,  
we find 
\bq
\lb{3.32}
\theta^{I}_{l} = -  \frac{|\alpha| \, \bar{v}}{rf^{2}},\;\;\;
\theta^{I}_{n} =  \alpha^2 \, \frac{|\alpha| \, \bar{u}} {rf^{2}} 
\eq
which are always non-positive in Region I.
Thus, all the symmetry spheres $\theta = $ const. are
trapped spheres (or circles if you prefer) 
in this region \cite{Gar00}.  Further, we can show from Eq.(\ref{3.29}) that
\bq
\lb{3.33}
R \equiv R_{ab}g^{ab} = 8\pi G \nabla_{a}\phi\nabla^{a}\phi =
\frac{2\alpha |\alpha|}{\left[ \alpha (-\bar{u}) + 
\bar{v}\right]^{4}},
\eq
which shows that, if $\alpha > 0$ the spacelike hypersurface $r(\bar{u},
\bar{v}) = 0$ is regular while if $\alpha < 0$ it is
singular. From the above expression we can also see that the
spacetime is free of singularity on the null hypersurface $\bar{v} = 0$, 
except for
the point $(\bar{u}, \, \bar{v} ) = (0, \, 0)$.  In order to have a 
geodesically maximal
spacetime, we must therefore extend the spacetime beyond the surface 
$\bar{v}=0$. 
There are, of course, many ways to make such an extension. However, 
we will only consider those that are analytic.  By imposing the condition 
of analyticity, 
the extension becomes unique and is given by simply
taking the corresponding solutions to be valid in the whole ($\bar{u}, 
\bar{v}$)-plane. In principle, one might obtain three extended regions in this 
way, $I',\; II,$ and $II'$, where  $ I'
\equiv \{x^{a}: \bar{u} \ge 0,\; \bar{v} \le 0\}$, $II
\equiv \{x^{a}: \bar{u} \le 0,\; \bar{v} \le 0\}$,   and $II' \equiv \{x^{a}: 
\bar{u} \ge 0,\; \bar{v}
\ge 0\}$ [{\it cf}. Fig.1]. 

\begin{figure}[htbp]
\begin{center}
\label{fig4}
\leavevmode
    \epsfig{file=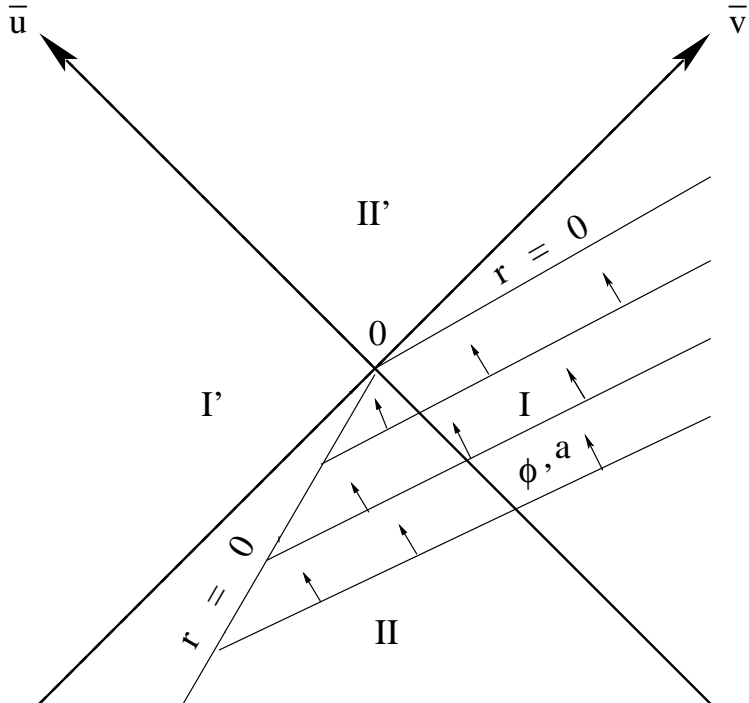,width=0.6\textwidth,angle=0}
\caption{The spacetime in the ($\bar{u}, \bar{v}$)-plane for the solutions 
described by Eqs.(\ref{3.29}) and (\ref{3.30}) with $\chi=1/2$. The 
hypersurfaces $\phi(\bar{u}, \bar{v}) = $ const. are the lines parallel 
to the one $r = 0$ in Region I, which is singular when $\alpha < 0$, 
and singularity-free when
$\alpha > 0$.  In constrast, note that the line $r = 0$ in Region
II is singular when $\alpha > 0$, and regular when $\alpha < 0$.  
For all values of $\alpha$, the spheres $\theta = $ const. are trapped in
Region I. In Regions I and II $\nabla^{a}\phi$ is timelike when $\alpha 
< 0$, and spacelike when $\alpha > 0$.}   
\end{center}    
 
\end{figure}

Applying this extension to the above solution, we find that the extended
solution in Region II is still given by  Eqs.(\ref{3.29}) and (\ref{3.30}),
with the corresponding physical quantities given by
Eqs.(\ref{3.31}--\ref{3.33}) but with $\bar{v} \le 0$ in these expressions. 
 From these equations we can see that the outgoing null geodesics are no
longer trapped in Region II and that the hypersurfaces 
$r(\bar{u}, \bar{v}) = $ const. become timelike.  Moreover, from the 
Ricci curvature it can be seen that the surface $r(\bar{u}, \bar{v}) = 0$ 
in this region is singular for $\alpha > 0$ while for $\alpha < 0$, it 
is regular.  In both cases, this surface will form the lower boundary 
of the extended region, such that the extended geodesically maximal 
spacetime in our case will consist of Regions I and II with the lines 
$r = 0$ as its upper and lower boundaries, as shown in Fig. 4.  The 
remaining portions of Fig. 4 are causally disconnected from the extended 
spacetime.  
Eq.(\ref{3.33}) also shows that $\nabla^{a}\phi$ is timelike only when
$\alpha < 0$. Thus, in order to interpret the extended solution as
representing gravitational collapse of the scalar field, 
we must take $\alpha < 0$, a condition which we shall assume from now
on for the case $\chi=1/2$.  The hypersurfaces 
$\phi(\bar{u}, \bar{v}) = $ const.
are thus straight lines parallel to the singular surface $r = 0$ in
Region I, as shown in Fig. 4.  The corresponding Penrose diagram is given
by Fig. 5.  From it, we can interpret Region I as the
interior of a black hole with the hypersurface $\bar{v} = 0$ (or 
$v = 0$) as its event horizon. 

In Region II, introducing the coordinates ${u}$ and
$v$ via the relations  
\bq
\lb{eq1}
\bar{u} = - \sqrt{-{u}},\;\;\; \bar{v} = - \sqrt{-v}, 
\eq
we can find the corresponding metric and massless scalar field in terms
of these variables.  Using a modified definition of $z\equiv v/u$, we have
\bqn
\lb{eq2}
ds^{2} &=& - 2e^{2\sigma(z)}du dv + (-u)^{2}s^{2}(z) d\theta^{2} \nb\\ 
\phi(u, v) &=& c\ln\left|-u\right| + \varphi(z) 
\eqn
where 
\bqn
\lb{eq3}
\sigma(z) &=& \ln\left|z^{1/4} + |\alpha| z^{-1/4}\right| + \sigma_0^1 \nb\\
s(z) &=& \alpha^{2} - z 
\nb\\
\varphi(z) &=& 2c\ln\left||\alpha| + z^{1/2}\right| +
\phi^{1}_{0}  \;\;\;\; ({\rm Region} \;II).
\eqn 
In the next section, when we study the perturbations of this $\chi = 1/2$
case, we will use this form of the metric as the background solution.

\begin{figure}[htbp]
\begin{center}
\label{fig5}
\leavevmode
    \epsfig{file=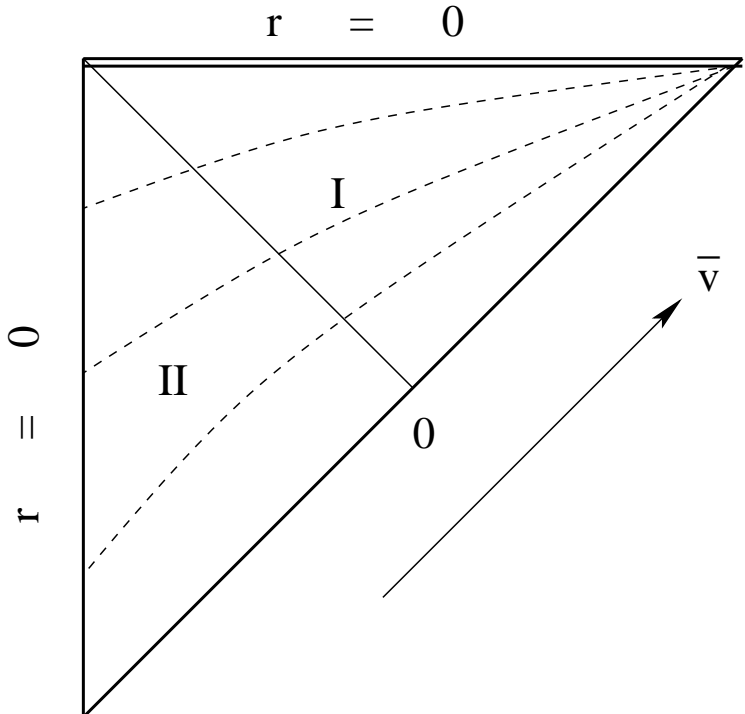,width=0.6\textwidth,angle=0}
\caption{This is the Penrose diagram for the solutions
described by Eqs.(\ref{3.29}) and (\ref{3.30}) with $\alpha < 0$ or,
equivalently, the Penrose diagram for the solutions described by  
Eqs.(\ref{3.35}) and (\ref{3.36}) with $\alpha > 0$.  The horizontal
line $r = 0$ is singular, while the vertical line $r = 0$ is free of spacetime
singularity. The dashed lines are the hypersurfaces where $\phi(\bar{u}, 
\bar{v}) =
$ const. The normal vector to these surfaces, $\nabla^{a}\phi$, is always 
timelike. In Region I all the surfaces $\theta = const $ are trapped, while
in Region II they are not.  }     
\end{center}      
\end{figure}  
 
Before moving on to consider the other cases, we note that, if $A(u, v)$ is a
solution of the Einstein  field equations
(\ref{2.7}--\ref{2.8}) with $\tilde{V} = 0$, then, the solutions
\bq
\lb{3.34}
A'(u, v) \equiv A(c_{1}u, c_{2}v),
\eq
also satisfy those equations, where $A =\{\sigma, \; r, \; \phi\}$, and
$c_{1}$ and $c_{2}$ are arbitrary constants. As an example, let us consider
the solution $A'(\bar{u},\bar{v}) = A(\bar{u}, - \bar{v})$, that is,
\bqn
\lb{3.35}
ds^{2} &=& - \frac{2}{|\alpha|} \, f^{2}(\bar{u},\bar{v}) d\bar{u}d\bar{v} 
           + r^{2}(\bar{u},\bar{v})d\theta^{2} \nb\\
r      &=& \alpha^{2}(-\bar{u})^{2} - \bar{v}^{2}  \nb\\
\phi   &=& 2c\ln\left|f(\bar{u},\bar{v})\right| + \phi^{1}_{0} 
\eqn
where  
\bq
\lb{3.36}
f(\bar{u},\bar{v}) \equiv   \alpha (- \bar{u}) - \bar{v}.
\eq
Clearly, this solution can be obtained from that of Eqs.(\ref{3.29}) and
(\ref{3.30}) by changing the sign of  $\alpha$. Thus, the solution of
Eqs.(\ref{3.35}) and (\ref{3.36}) with $\alpha > 0$ also represents
gravitational collapse of the massless scalar field, and the corresponding
Penrose diagram is the same as that given by Fig. 5.
It can be shown that in terms of ${u}$ and  $v$ defined by
Eq.(\ref{eq1}) the metric and the massless scalar field in the present case
are also given by Eqs.(\ref{eq2}) and (\ref{eq3}) but now with $\alpha > 0$.

\subsubsection{ $ \frac{1}{2} < \chi < 1$}

To begin analyzing Eqs.(\ref{3.25}--\ref{3.27}) for 
the case that $1/2 < \chi < 1$, we again introduce new coordinates
$\bar{u}$ and $\bar{v}$ via the relations 
\bq
\lb{3.37}
\bar{u} = - (-{u})^{1/2n},\;\;\;\;\;
\bar{v}= v^{1/2n} 
\eq
where we have defined the constant 
\bq
n \equiv \frac{1}{2(1 - \chi)} > 1 
\eq
(and which will shortly be taken to be an
integer, but can be assumed, for the moment, to be real).  
In these coordinates, the metric and the massless scalar field are given by
\bqn
\lb{3.38}
ds^{2} & = & - \frac{8 n^{2}}{(16\alpha^2)^\chi}  \, 
                 \left|f(\bar{u},\bar{v})\right|^{4\chi} \, d\bar{u}d\bar{v} 
             + r^{2}(\bar{u},\bar{v})d\theta^{2}   \nb\\ 
  r    & = &   \alpha^{2}(- \bar{u})^{2n} 
             - \bar{v}^{2n}   \nb\\
 \phi  & = &   2c\ln\left| f(\bar{u},\bar{v}) \right| 
             + \phi^{1}_{0} 
\eqn
where  
\bq
\lb{3.39}
f(\bar{u},\bar{v}) \equiv   \alpha (-\bar{u})^{n} + \bar{v}^{n}. 
\eq
The relevant physical quantities are given by
\bqn
\lb{3.40}
\theta^{I}_{l} & = & - { (16\alpha^2)^\chi \over 4n } \,  
                      { \bar{v}^{2n-1} \over r \, |f(\bar{u},\bar{v})|^{4\chi}} 
\nb\\
\theta^{I}_{n} & = & - { \alpha^{2} \, (16\alpha^2)^\chi \over 4n } \, 
                      { (-\bar{u})^{2n-1} \over r \, |f(\bar{u},\bar{v})|^{4\chi}}   
\nb\\
\nabla_{a}r\nabla^{a}r & = & - (16\alpha^2)^\chi  
                               { (-\bar{u}\bar{v})^{n-1} 
                                        \over  
                                 \left|   \alpha (-\bar{u})^{n} 
                                        + \bar{v}^{n}          \right|^{4\chi}} 
\nb\\
 R & = &  8\pi G \nabla_{a}\phi\nabla^{a}\phi 
     =  2\alpha \chi (16\alpha^2)^\chi \,  
        { (-\bar{u}\bar{v})^{n-1} 
                 \over 
          \left|\alpha (-\bar{u})^{n} + \bar{v}^{n}\right|^{2(2\chi + 1)}}.
\eqn
Because we have $n > 1$, we can see from the above expressions that all 
the surfaces of symmetry, $\theta = $ const. are trapped  in
Region I and that the spacetime is bounded from above by the hypersurface 
$\bar{v} = |\alpha|^{1/n}\bigl( - \bar{u}\bigr)$ ({\it i.e.} $\; r = 0$), 
which is singular for $\alpha < 0$, and regular for $\alpha > 0$.  Note 
that as $\bar{v} \rightarrow 0^{+}$, we have $R \rightarrow 0$ and the 
spacetime is perfectly regular on $\bar{v} = 0$ as well.  Therefore, in 
order to have a geodesically maximal spacetime, we must extend the spacetime 
beyond this surface to $\bar{v}<0$.  As in the previous case 
with $\chi = 1/2$, we can
simply take the above solutions to also be valid in Region II.  However, this
extension will not be analytic unless $n$ is an integer.  When $n$ is not an
integer, the extended metric will not even be real.  As an example, consider the
case where $n$ is rational, {\it i.e.} $n = (2m+1)/(2l)$, where $l$ and $m$ 
are integers.  A possible extension in this case would be
\bq
\lb{3.41}
A^{II}(\bar{u}, \bar{v}) = A^{I}(\bar{u}, -\bar{v}).
\eq
Clearly, this extension is not analytic, too, but it guarantees that the
metric in Region II is real and the extension has the maximal order of
derivatives in comparing with any of other extensions. 

However, as mentioned previously, we will only consider those
cases where the extensions are analytic.  Therefore, we shall
restrict ourselves in the following to those cases where $n$ is an 
integer.  For such an extension, the relevant physical quantities in 
Region II will also be given by Eq.(\ref{3.40}) but now with 
$\bar{v}\le0$.  With this, we find that in Region II, in order to have 
$\nabla^{a}\phi$ be timelike we must have 
\bq
\lb{3.42}
n = \cases{2l, & $\alpha > 0$,\cr
2l +1, & $\alpha < 0$,\cr}
\eq
where $l$ is an integer. To further study these maximally extended solutions, 
let us consider the cases $\alpha > 0$ and $\alpha < 0$, separately.

\bigskip

\noindent {\bf Case (a) $\; \alpha > 0$}:  In this case we must choose $n = 2l$ so 
that
$\nabla^{a}\phi$ is timelike in Region II, and the corresponding solutions 
can
be interpreted as representing gravitational collapse in this region.    
$\nabla^{a}\phi$ is spacelike in Region I,  while on the hypersurface 
$\bar{v} = 
0$ it
becomes null, as we can see from Eq.(\ref{3.40}). The hypersurfaces
$\phi(\bar{u}, \bar{v}) =$ const. are shown in Fig. 6. Similar to  the last 
case, we may
interpret Region I as the interior of a black hole, since  the
surfaces $\theta = $ const. are trapped, although  
no spacetime singularity is present  in this case [except for the point
$(\bar{u}, \bar{v}) = (0, 0)$]. This black hole can be considered as the final 
state of
the collapse of the massless scalar field in Region II. This can be seen 
most clearly
from an analysis of the energy-momentum tensor. 
Near the null hypersurface, $\bar{v} = 0$, the only non-vanishing 
component of
the energy-momentum tensor is given by  
\bq
\lb{3.43} 
T_{\bar{u}\bar{u}} = \frac{4\chi n^{2}}{(-\bar{u})^{2}}, 
\eq 
which represents a pure energy flow  along the null hypersurface $\bar{u} =
$ const. moving from Region II into Region I.  The corresponding Penrose
diagram is also given by Fig. 5, but now both of the vertical and horizontal
lines $r = 0$ are free of spacetime singularities, except for the point 
$(\bar{u},
\bar{v}) = (0, 0)$.

\begin{figure}[htbp]
\begin{center}
\label{fig6}
\leavevmode
    \epsfig{file=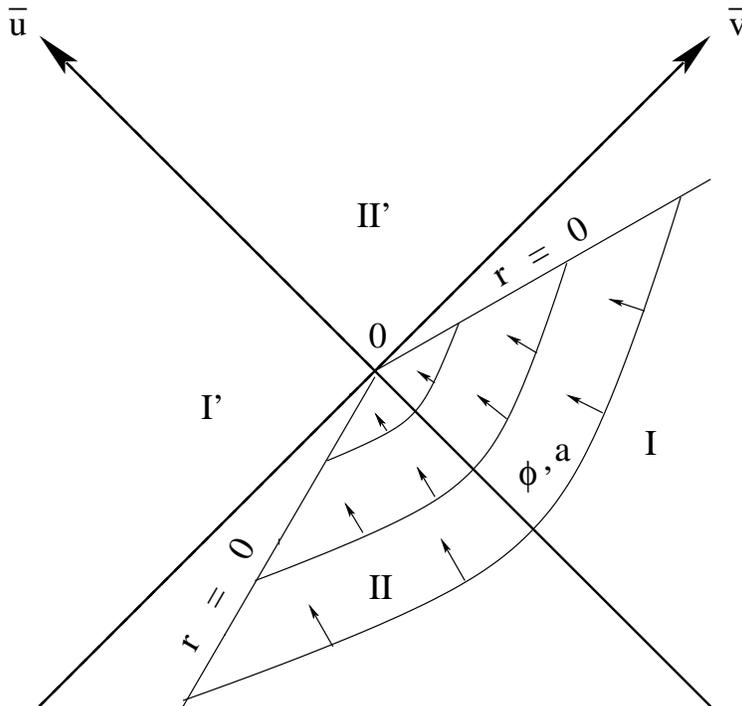,width=0.6\textwidth,angle=0}
\caption{The hypersurfaces $\phi(\bar{u},\bar{v}) = $ const. in 
the ($\bar{u},\bar{v}$)-plane for the spacetime described by Eqs.(\ref{3.38}--\ref{3.40},\ref{metric4}--\ref{eqa2}) with $\alpha > 0$ and 
$n = 2l \; (\alpha < 0$ and $ n = 2l+1)$, where $l$ is an integer.  
The normal vector to these surfaces $\nabla^{a}\phi$ is timelike in 
Region II and
null on the hypersurface $v = 0$ for both of the two cases $\alpha > 0$ and
$\alpha < 0$, while in Region I it is spacelike  for $\alpha > 0$ and
timelike for $\alpha < 0$.  For both cases, the spheres $\theta = $ const. 
are trapped in
Region I but not in Region II.  The spacetime is free of singularities on
the surfaces $r = 0$ in both regions when $\alpha > 0$, while 
for $\alpha < 0$, the surface $r = 0$ in Region I is singular.  The sole
exception to this is 
the point $(\bar{u}, \bar{v}) = (0, 0)$ which is always singular. }      
\end{center}
\end{figure}  

In Region II, introducing ${u}$ and ${v}$ via the relations,
\bq
\lb{eqa1}
u = - (-\bar{u})^{4l},\;\;\;\;
v = - (- \bar{v})^{4l},\;\;\; 
\left(\alpha > 0,\; {\rm{region}}\; II\right), 
\eq
we find that the metric and the massless scalar field in terms of $u$
and $v$ (again using the modified definition of $z\equiv v/u$) are given by
\bqn
\lb{metric4}
ds^{2} &=& -2e^{2\sigma(z)} du dv + (-u)^{2}s^{2}(z) d\theta^{2},\nb\\
\phi(u,v) &=& c\ln|-u| + \varphi(z),
\eqn
where
\bqn
\lb{eqa2}
\sigma(z) &=& 2\chi\ln\left|z^{1/4} + |\alpha| z^{-1/4}\right| 
+ \sigma^{1}_{0} \nb\\ 
s(z) &=& \alpha^{2} - z \nb\\
\varphi(z) &=& 2c\ln\left||\alpha| + z^{1/2}\right| +
\phi^{1}_{0},\;\;\; \left({\rm{region}}\;  II\right).
\eqn 
In the next section, when we study perturbations we shall use this form for  
the metric as the background spacetime for this case.

\bigskip

\noindent {\bf Case (b) $\; \alpha < 0$}:  In this case we must choose $n = 2l + 1$ 
so that $\nabla^{a}\phi$ is timelike in Region II. This region is bounded 
by the timelike hypersurface $r = 0$, on which the spacetime is regular.  
Unlike the case $\alpha > 0$, in Region I now $\nabla^{a}\phi$ is still 
timelike,  while on the hypersurface $\bar{v} = 0$ it is null. The 
hypersurfaces $\phi(\bar{u}, \bar{v}) = $ const. are also 
given in Fig. 6. However, now the spacelike hypersurface $r
= 0$ in Region I is singular and the surfaces $\theta = $ const.
are trapped surfaces. Near the null hypersurface, $\bar{v} = 0$, the
energy-momentum tensor also has only one non-vanishing component, given
exactly by Eq.(\ref{3.43}). Therefore, in the present case the corresponding
solutions can also be interpreted as representing gravitational collapse of a
massless scalar field, with a black hole finally formed in Region I.
The corresponding Penrose diagram is again that of Fig. 5.  

In Region II, on defining ${u}$ and $v$ as
\bq
\lb{eqa3}
u = - (-\bar{u})^{2(2l + 1)},\;\;\;\;
v = - (-\bar{v})^{2(2l + 1)},\;\;\; 
\left(\alpha < 0,\; {\rm{region}}\;  II\right),
\eq
we find that in terms of ${u}$ and $v$ the metric
and the massless scalar field are exactly given by Eqs.(\ref{metric4}) and
(\ref{eqa2}).

\subsubsection{$\chi \ge 1$}

In this case it can be shown that the
spacetime is already geodesically maximal in the region $v \ge 0,\; u \le 0$,
and does not need to be extended beyond the hypersurface $v = 0$ \cite{CF01}. 
Indeed, it is found that the null geodesics $u = $ const. have the integral  
\bq
\lb{3.44}
\eta = \cases{\eta_{0}v^{1-\chi}, & $\chi > 1$,\cr
\beta\ln(v), & $\chi = 1$,\cr}
\eq
near the hypersurface $v = 0$, where $\eta$ denotes the affine
parameter along the null geodesics, and $\beta$ and $\eta_{0}$ are
integration constants. Thus, as $v \rightarrow 0^{+}$, we always have
$\eta \rightarrow \pm \infty$ for $\chi \ge 1$. On the other hand, from
Eq.(\ref{3.27}) we can see that the spacetime is singular on the spacelike
hypersurface $v = - \alpha u$ (or $\; r = 0$) for $\alpha < 0$, and 
regular on this surface for $\alpha > 0$.  Moreover, within the entire 
region, the symmetry spheres $\theta = $ const. are trapped surfaces, as
can be seen from the expansions $\theta_{l}$ and $\theta_{n}$ which 
are always non-positive  
\bq
\lb{3.45}
\theta_{l} = - \frac{1}{r}e^{-2\sigma},\;\;\;\;
\theta_{n} = - \frac{\alpha^{2}}{r}e^{-2\sigma},\;\; (u \le 0,\; v \ge 0).
\eq
It is therefore difficult to interpret the solutions in the present case 
as representing gravitational collapse. 

\begin{figure}[htbp]
\begin{center}
\label{fig7}
\leavevmode
    \epsfig{file=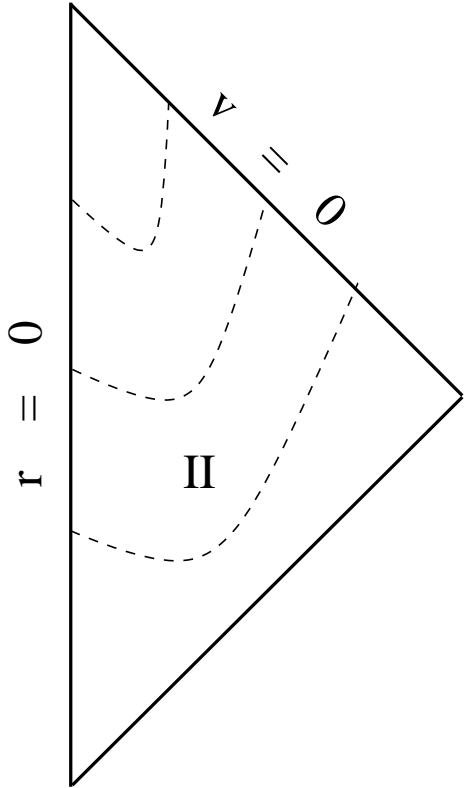,width=0.4\textwidth,angle=0}
\caption{The Penrose diagram for the solutions given by
Eqs.(\ref{3.47}). The dashed lines represent the
hypersurfaces   $\phi(u,v) = $ const., the normal vector of which is always
timelike.   The whole spacetime is free of singularity   except for the point
($u, v) = (0, 0$). }
\end{center}

\end{figure}

However, applying Eq.(\ref{3.41}) to the above solutions, we can consider
the solutions in Region II, but which are now causally disconnected 
from Region I.  These solutions are given by  
\bqn 
\lb{3.46}  
e^{2\sigma(u, v)} &=& \frac{1}{(16\alpha^2)^\chi} \, 
                      \frac{\left[\alpha (-u)^{1/2} + (-v)^{1/2}\right]^{4\chi}}
                           {(u v)^{\chi}},
\nb\\ 
r(u,v) &=& \alpha^{2}(-u) + v,\nb\\ 
\phi(u,v)&=& 2c\ln\left|\alpha(-u)^{1/2} + (-v)^{1/2}\right|
+ \phi^{1}_{0},\;\;\;\;
(u,\; v \le 0). 
\eqn
 From these expressions, we find that
\bq 
\lb{3.48}
\theta_{l} = - \alpha^{-2} \, \theta_{n} = \frac{e^{-2\sigma}}{r},
\eq
which is non-negative; that is, the outgoing null geodesics are now no longer
trapped.  One can show that the null hypersurface $v = 0$ is also 
future null infinity of Region II.  Thus, the spacetime is
geodesically maximal in this region. As a result, we also have 
\bq
\lb{3.47}
R = 8\pi G \nabla_{a}\phi\nabla^{a}\phi
  =  - 2 \alpha \chi (16\alpha^2)^\chi \, \frac{(uv)^{ \chi -1/2} }
       {\left|(-v)^{1/2} + \alpha (-u)^{1/2}\right|^{2(2\chi + 1)}}, 
\eq
which shows that, when $\alpha < 0$ the spacetime in this region
is singular at $r = 0$, and when $\alpha > 0$ the spacetime is free of 
such  singularities. Moreover,  from Eq.(\ref{3.47})  we can also see that 
the normal vector to the  hypersurfaces $\phi(u, v) = $ const. is always 
timelike when $\alpha > 0$, and quite similar to those given by Fig. 6. Thus,
the solutions with $\alpha > 0$ can be thought of as representing
gravitational collapse starting from past null infinity at $u = -\infty$.
Thus, we will only consider solutions with $\alpha > 0$ for the current
case of $\chi \ge 1$.  As a result, the spacetime is free of singularities 
at the
beginning of the collapse, but with a spacetime singularity forming 
at the point $(u,\, v) = (0,\, 0)$. The corresponding Penrose diagram is given
in Fig. 7.  It is interesting to note that in this case, although no 
event horizon is formed, the spacetime singularity at the point 
$(u, v) = (0, 0)$ is not naked and the only observers who are able to see
the singularity there are those observers who arrive at that point.

It can be shown that in this case (with $\alpha > 0$) the metric and the
massless scalar field in Region II can be also written in the same form as
that given by Eqs.(\ref{metric4}) and (\ref{eqa2}). 

In fact, from   Eqs.(\ref{eq2}--\ref{eq3}) and 
Eqs.(\ref{metric4}--\ref{eqa2}), we find that {\em in the case $a_{0}b_{0} 
\not=0$, all the solutions that represent gravitational collapse of the 
massless scalar field can be cast in the form of  Eqs.(\ref{metric4}) and 
(\ref{eqa2}) in Region II}.  For that reason, when we consider linear 
perturbations in the next section, we will always refer to 
Eqs.(\ref{metric4}) and (\ref{eqa2}) with $\chi \ge 1/2$ when 
considering the case of $a_{0}b_{0} \not= 0$.  In addition, because it is
only the absolute value of $\alpha$ which enters these equations, we will
from now on and without loss of generality assume that $\alpha > 0$.

\section{Linear Perturbations of the Self-Similar Solutions}

We now turn to a study of the linear perturbations of the self-similar solutions
presented in the last section.  Let us consider perturbations given by
\bqn
\lb{4.1}
\sigma(z, \tau) &=& \sigma_{0}(z) + \epsilon \, \sigma_{1}(z)e^{k\tau},\nb\\
s(z, \tau) &=& s_{0}(z) +  \epsilon \, s_{1}(z)e^{k\tau},\nb\\
\varphi(z, \tau) &=& \varphi_{0}(z) +  \epsilon \, \varphi_{1}(z)e^{k\tau},
\eqn
where $\epsilon$ is a very small real constant, and quantities with subscripts
``1" denote perturbations, and those with
``0" denote the self-similar solutions, given, respectively, by
Eq.(\ref{3.8}) with $s_{0} = z,\; \sigma^{1}_{0} = 0$, Eq.(\ref{3.16}) with
$s_{0} = 1, \; \sigma^{1}_{0} = 0$,  and  Eqs.(\ref{metric4}) and
(\ref{eqa2}) with $\chi \ge 1/2$. 
It is understood that there may be many perturbation modes for different
values (possibly complex) of the constant $k$.  The general perturbation 
will be the sum of these individual modes.  Those modes with $Re(k) > 0$ 
grow as $\tau \rightarrow
\infty$ and are referred to as unstable modes, while the ones with
$Re(k) < 0$ decay and are referred to as stable modes.  By definition, 
critical solutions will have one and only one unstable mode.  It should be noted
that in writing Eq.(\ref{4.1}), we have already used some of the gauge 
freedom available to us to write  
the perturbations such that they preserve the form of the metric (\ref{2.6}). 
However, this does not completely fix the gauge freedom and we will return
to this issue when we consider gauge modes.

To first order in $\epsilon$, it can be shown that the non-vanishing
components of the Ricci tensor are given by 
\bqn 
\lb{4.2} 
R_{uu} & = & - \frac{1}{r}\left(r_{,uu} - 2\sigma_{,u}r_{,u}\right)
\nb\\ 
       & = & - \frac{e^{2\tau}}{u^{2}_{0} s_{0}}  
                  \left[   z^{2}s''_{0} 
                         - 2z\sigma'_{0}\left(zs'_{0} - s_{0}\right)
                 \right]  
\nb\\
  & & \quad  - \epsilon \, \frac{ e^{(2+k)\tau}}{u^{2}_{0} s_{0} }
                  \Bigg\{   z^{2}s''_{1} 
                          + 2z(k - z\sigma'_{0})s'_{1}
\nb\\ 
  & & \qquad\qquad\qquad  - \left[   z^{2} s^{-1}_{0} \left(  s''_{0} 
                                                            - 2\sigma'_{0}s'_{0}
                                                     \right) 
                                   + k ( 2z \sigma'_{0} + 1 - k )
                           \right] \, s_{1} 
\nb\\
  & & \qquad\qquad\qquad  - 2 ( zs'_{0} - s_{0} ) 
                              ( z\sigma'_{1} + k \sigma_{1}) \; 
                  \Bigg\} 
\nb\\  
R_{vv} & = & - \frac{1}{r}\left(r_{,vv} - 2\sigma_{,v}r_{,v}\right)
\nb\\
       & = & - \frac{e^{2\tau}}{u^{2}_{0} s_{0}} 
                  \left(s''_{0} - 2\sigma'_{0}s'_{0}\right)
\nb\\
  & & \quad  - \epsilon \, \frac{e^{(2+k)\tau}}{u^{2}_{0} s_{0}}
                  \Bigg\{   s''_{1} 
                          - 2\sigma'_{0}s'_{1} 
                          - s^{-1}_{0}\left(   s''_{0} 
                                             - 2\sigma'_{0}s'_{0}
                                     \right) \, s_{1} 
                          - 2s'_{0} \sigma'_{1} 
                  \Bigg\} 
\nb\\
R_{uv} & = & - \frac{1}{r}\left(r_{,uv} + 2r\sigma_{,uv}\right)
\nb\\
       & = & - \frac{e^{2\tau}}{u^{2}_{0} s_{0}} 
                  \Big[   zs''_{0} 
                         + 2s_{0} \left(z\sigma''_{0} + \sigma'_{0}\right)
                  \Big]
\nb\\
  & & \quad  - \epsilon \, \frac{e^{(2+k)\tau}}{u^{2}_{0} s_{0}}
                 \Bigg\{   zs''_{1} 
                         + 2zs_{0}\sigma''_{1} 
                         + k s'_{1} 
                         + 2(1 + k) s_{0}\sigma'_{1} 
                         - z s^{-1}_{0} s''_{0} s_{1}
                 \Bigg\} 
\nb\\
R_{\theta\theta} 
       & = &   2 e^{-2\sigma} r r_{,uv} 
\nb\\
       & = &   2 z s_{0} s''_{0} e^{-2\sigma_{0}}
\nb\\
  & & \quad  + \epsilon \, 2 e^{k\tau - 2\sigma_{0}}s_{0}
                 \bigg\{   zs''_{1} 
                         + ks'_{1} 
                         + zs''_{0}\left(   s^{-1}_{0}s_{1} 
                                          - 2\sigma_{1}
                                   \right)
                 \bigg\}.
\eqn 
To zeroth order the Einstein field equations are given by
Eqs.(\ref{2.13a}--\ref{2.13d}), while to first order they take the form
\bqn 
\lb{4.3a}
        z^{2}s''_{1} 
      + 2z(k - z\sigma'_{0})s'_{1}
\qquad\qquad\qquad\qquad\qquad\qquad
 & &  
\nb\\ 
      - \left[   z^{2}s^{-1}_{0}\left(s''_{0} - 2\sigma'_{0}s'_{0}\right)
               + k(2z\sigma'_{0} + 1 -k)
       \right] \, s_{1} 
 & &
\nb\\
      - \, 2 ( zs'_{0} - s_{0} )
          ( z\sigma'_{1} + k \sigma_{1}) 
\qquad\qquad\qquad\qquad
 &=&  - 16\pi G s_{0} \left( z\varphi'_{0} - c \right) 
                      \left( z\varphi'_{1} + k \varphi_{1} \right) 
   \\
\lb{4.3b}
        s''_{1} 
      - 2\sigma'_{0}s'_{1} 
      - s^{-1}_{0}\left(   s''_{0} 
                         - 2\sigma'_{0}s'_{0}
                 \right) \, s_{1} 
      - 2s'_{0} \sigma'_{1} 
 &=&  - 16\pi G s_{0} \varphi'_{0} \varphi'_{1} 
   \\
\lb{4.3c}
        zs''_{1} 
      + 2zs_{0}\sigma''_{1} 
      + k s'_{1}	\qquad\qquad\qquad
 & &
\nb\\
      + \, 2(1 + k)s_{0}\sigma'_{1}    
      - z s^{-1}_{0}s''_{0}s_{1}
 &=&  - 8\pi G s_{0}\left[   \left(2z\varphi'_{0} - c\right)\varphi'_{1} 
                           + k \varphi'_{0}\varphi_{1}
                   \right] 
   \\
\lb{4.3d}
        zs''_{1} 
      + ks'_{1} 
      + zs''_{0}\left(   s^{-1}_{0} s_{1} 
                       - 2 \sigma_{1}
               \right) 
 &=&    0. 
\eqn
To zeroth order, the equation for the massless scalar field is given 
by Eq.(\ref{2.13e}) while, to first order, the scalar field equation 
is found to be 
\bqn
\lb{4.4}
         2zs_{0}\varphi''_{1} 
      + \left[  2zs'_{0}  +  (1 + 2k)s_{0} \right] \, \varphi'_{1} 
      + ks'_{0}\varphi_{1} \qquad\qquad 
 & & 
\nb\\
      + \left(2z\varphi'_{0} -c\right)s'_{1} 
      + \left[2z\varphi''_{0} + (1+k)\varphi'_{0}\right] \, s_{1} 
 &=&    0.
\eqn

In order to be somewhat systematic in our examination of the perturbation
equations, we will consider separately the same three cases that we did in 
section III, 
namely $a_{0}\not= 0,\; b_{0} = 0;\; a_{0} = 0,\; b_{0} \not =  0$ and 
$a_{0} b_{0} \not= 0$.  The notation will of necessity become rather 
complicated and we ask for the reader's indulgence.  For our part, we will try
to keep the complexity under control and the physical interpretation near
the fore.

\subsection{$a_{0} \not= 0$ and $ b_{0} = 0$}

In this case, the self-similar background solutions are given by Eq.(\ref{3.8})
with $s_{0}(z) = z$ and $\sigma^{1}_{0} = 0$. Substituting these solutions 
into
Eqs.(\ref{4.3a}--\ref{4.4}) we find 
\bqn
\lb{4.5a}
  zs''_{1} + k s'_{1} & = & 0 
\\
\lb{4.5b}
  (k - \chi) \left[   zs'_{1} 
                    + (k - 1)s_{1}
            \right] & = & 0 
\\
\lb{4.5c}
  z(k + \chi) s'_{1} - \chi s_{1} + 2z^{2}\sigma'_{1} 
     & = & 16\pi c G \, z^{2}\varphi'_{1} 
\\
\lb{4.5d}
  z^{2}\sigma''_{1} + (1+k) z \sigma'_{1} 
     & = & - 4\pi c G \, (k\varphi_{1} + z\varphi'_{1}) 
\\
\lb{4.5e}
  2 z^{2}\varphi''_{1} + (3+2k)z\varphi'_{1} + k\varphi_{1} 
     & = & - \frac{c}{z}\left[zs'_{1} + (k -1)s_{1}\right].
\eqn
It should be noted that in writing Eqs.(\ref{4.5b}--\ref{4.5e}), we have used
Eq.(\ref{4.5a}). As usual, Eqs.(\ref{4.5b}) and (\ref{4.5c}) are constraints, 
and Eqs.(\ref{4.5a}), (\ref{4.5d}) and (\ref{4.5e}) are the dynamical 
equations.  Due to the redundancy in the Einstein equations, these are not
all independent and, {\it e.g.} it can be shown that Eq.(\ref{4.5d}) can be 
written in terms of the others.  

Once these equations are integrated, one needs boundary conditions to 
complete the solution.  Such boundary conditions are crucial for 
a proper determination of the eigenmodes but are rather subtle in general 
relativity.  Indeed, there is no fixed prescription to follow 
\cite{Gar00,CF01}. In the present case, the background spacetime has four 
boundaries, $u = \pm \infty$ and $v 
= 0, \; \infty$. However, since the perturbation equations are of second 
order, it is usually sufficient to impose boundary conditions on only two 
hypersurfaces. As $\sigma_{1},\; s_{1}$ and $\varphi_{1}$ are functions of $z$ 
only, it is rather natural to choose the hypersurface $z = z_{0}$ as one of 
the two boundaries, that is, we 
assume that the perturbations are turned on at $z = z_{0}$, where $z_{0}$ is 
an arbitrary constant. Prior to $z = z_{0}$ the perturbations are zero. 
In order to 
match the perturbations smoothly across this surface, we require that 
$s_{1}(z_{0}) = 0 = \varphi_{1}(z_{0})$. Also, by using the coordinate 
tranformations (\ref{2.6aa}) we can always set $\sigma_{1}(z_{0}) = 0$. Then, 
on the hypersurface $z = z_{0}$ we have the boundary conditions,
\bq
\lb{4.5f}
\sigma_{1}(z_{0}) = s_{1}(z_{0}) = \varphi_{1}(z_{0}) = 0.
\eq
The second boundary has to be one of the two hypersurfaces, 
$v = 0$ and $u = \infty$.  As shown in the last section, the 
hypersurface $v = 0$ is always singular in the present case 
and we do not know how to 
impose boundary conditions on such a surface.\footnote{Sometimes the 
condition is imposed  
that the perturbations should be ``less singular" than the 
background. However, it is not always clear if the expression ``less 
singular" should refer to the spacetime curvature or to the metric 
components.}  Therefore, in the 
following we shall choose the hypersurface $u = \infty$ as the second 
boundary.  Because this surface represents future null infinity for this  
spacetime, it is reasonable to assume that the perturbations grow slower than 
the background.  In the present case, we have $s_{0}(z) \sim z,\; 
\sigma_{0}(z) \sim \ln (z), \; \varphi_{0}(z) \sim \ln (z)$. However, 
considering the fact 
that $e^{\tau} \sim (-u)^{-1}$, we might require that, as 
$u \rightarrow \infty$, $\delta r(u,v) \equiv 
\epsilon (-u) s_{1}(z) e^{k\tau}$ and $\delta \varphi(u,v) \equiv \epsilon 
\varphi_{1}(z)e^{k\tau}$
be finite while $\delta \sigma(u,v) \equiv \epsilon \sigma_{1}(z)e^{k\tau}$ 
is requied to grow slower than $\ln(-u)$, as $u \rightarrow \infty$.  
Instead, we will choose to impose somewhat stronger conditions here, namely
\bq
\lb{4.5g}
\delta r(u,v),\;\; \delta \varphi(u,v),\;\;  \delta \sigma(u,v)   
\rightarrow\;\;  {\rm{finite}}, 
\eq
as $u \rightarrow \infty$.  Clearly, if we find unstable modes with this  
stricter criterion of regularity, then in general the solutions will still be   
unstable.

Before proceeding to the full perturbation analysis, we need to address  
questions concerning gauge modes.  Under the coordinate transformations 
\bq
\lb{4.5ga}
u  \rightarrow u + \epsilon \xi_{1}(u),\;\;\;\;
v  \rightarrow v + \epsilon \xi_{2}(v),
\eq
we generate perturbations of the form 
\bqn
\lb{4.5gb}
\delta\sigma &=& \frac{\epsilon}{2}\left\{ \chi \left[\frac{\xi_{2}(v)}{v} - 
\frac{\xi_{1}(u)}{u}\right] + \xi'_{1}(u)  + \xi'_{2}(v)\right\} \nb\\
\delta r &=& \epsilon\xi_{2}(v) \nb\\
\delta\varphi &=& \epsilon c   \left(\frac{\xi_{2}(v)}{v} - 
\frac{\xi_{1}(u)}{u}\right),
\eqn
where $\xi_{1}(u)$ and $\xi_{2}(v)$ are arbitrary functions.
For such perturbations to take the forms of Eq.(\ref{4.1}), we must 
choose
\bq
\lb{4.5gc}
\xi_{1}(u) = \frac{c_{1}}{c}{u_{0}}^{k}(-u)^{1-k},\;\;\;\;
\xi_{2}(v) = c_{2} {u_{0}}^{k} v^{1-k} 
\eq
where $c_{1}$ and $c_{2}$ are arbitrary constants. Then, the solutions 
to the  
perturbation equations due to the gauge transformations are given by
\bqn
\lb{4.5gd}
s_{1}(z) &=& c_{2} z^{1-k} \nb\\
\sigma_{1}(z) &=& \frac{c_2}{2} \, (1 - k + \chi)z^{-k} 
- \frac{c_{1}}{2c}(1- k - \chi)  \nb\\
\varphi_{1}(z) &=& c_{1} + cc_{2} \, z^{-k}.
\eqn 
It can now be seen that our boundary conditions, Eqs.(\ref{4.5f}) and 
(\ref{4.5g}) eliminate each of these gauge modes.

Now let us turn to solving the perturbation equations given by 
Eqs.(\ref{4.5a}--\ref{4.5e}).
From Eq.(\ref{4.5b}) we can see that there are two subcases which we need
to consider separately, $k = \chi$ and $k \not= \chi$. 

\subsubsection{$k = \chi$}

In this subcase, we have
\bq
\lb{4.6}
k = \chi \equiv 8\pi  G c^{2} > 0,
\eq
that is, the constant $k$ is real and strictly positive. Integrating
Eq.(\ref{4.5a}) we obtain
\bq
\lb{4.7}
s_{1}(z) = \cases{\beta \ln |z|    + s^{0}_{1}, & $ k = \chi     = 1$, \cr
                  \beta z^{1-\chi} + s^{0}_{1}, & $ k = \chi \not= 1$, \cr}
\eq
where $\beta$ and $s^{0}_{1}$ are real integration constants.

\bigskip

\noindent {\bf Case (a) $\; k = \chi = 1$}:  From 
Eqs.(\ref{4.5a}--\ref{4.5e}) we find that the solutions to the perturbation
equations can be written 
\bqn
s_{1}(z) & = &   \beta \ln |z| + s^{0}_{1}   
\nb\\
\sigma_{1}(z) & = &   8\pi cG \, \varphi_{1}(z) 
                    + {\beta - s^{0}_{1} \over 2z } 
                    - \frac{\beta}{2z} \, \ln |z| + \sigma^{0}_{1} 
\nb\\
\varphi_{1}(z) & = & c_{1} z^{-1/2} + c_{2} z^{-1} + \frac{c\beta}{z}\ln |z|  
\eqn
where $c_1$, $c_2$ and $\sigma^{0}_{1}$ are additional real, integration 
constants.
Applying the boundary conditions at $z=z_0$, Eq.(\ref{4.5f}), 
to these solutions, we find that
\bqn
s^{0}_{1} & = &  - \beta\ln z_{0}  
\nb\\ 
\sigma^{0}_{1} & = & - \frac{\beta}{2 z_0} 
\nb\\
c_{1} &=& - z_{0}^{-1/2}\left[c_{2} + c\beta \ln (z_{0})\right],
\eqn
from which we obtain 
\bqn
r(u,v) & = & (- u) \left[s_{0}(z) + \epsilon s_{1}(z) e^{k\tau}\right]\nb\\
       & = & v + \epsilon \beta u_{0}\ln\left(\frac{v}{- z_{0} u}\right).
\eqn
From this form for $r(u,v)$, we can see that the condition of finiteness 
as $u \rightarrow \infty$, Eq.(\ref{4.5g}), 
leads to $\beta = 0$. Setting $\beta = 
0$ in these solutions yields 
that $ \delta\sigma$ and $\delta\varphi$ are all finite as $u \rightarrow 
\infty$.  Therefore, in this case ($k=\chi$), there is a perturbation mode
with $k=1$ which 
satisfies the appropriate boundary conditions, Eqs.(\ref{4.5f}) 
and (\ref{4.5g}).  Of course, because $k > 0$, this is an unstable mode.
 
\bigskip

\noindent {\bf Case (b) $\; k = \chi \not= 1$}:  In this case, the perturbation equations
given by Eqs.(\ref{4.5a}--\ref{4.5e}) can again be integrated to yield
\bqn
s_{1}(z) & = & \beta  z^{1-\chi} + s^{0}_{1}
\nb\\
\sigma_{1}(z) & = &   8\pi cG \, \varphi_{1}(z) 
                    - \frac{\beta}{2} (2\chi - 1) \, z^{-\chi}  
                    - \frac{s^{0}_{1} \, \chi}{2} \, z^{-1} 
                    + \sigma^{0}_{1} 
\nb\\
\varphi_{1}(z) &=&\cases{   c_{1} z^{-1/2} 
                          + c_{2} z^{-\chi} 
                          + c \, s^{0}_{1} \, z^{-1}, & $k \neq 1/2$ , \cr 
                            c_{1} z^{-1/2} 
                          + c_{2} z^{-1/2} \, \ln |z|  
                          + c \, s^{0}_{1} \, z^{-1}, & $k = 1/2$ . \cr } 
\eqn
On application of the boundary conditions, Eqs. (\ref{4.5f}) and
(\ref{4.5g}), at $z=z_0$ and $u\rightarrow\infty$, it can be shown that 
there are nontrivial perturbations for $k > 1/2$ while for $k\le 1/2$,
no solutions exist which are compatible with the boundary conditions.   

We conclude this section by noting that all the solutions given by Eq.(\ref{3.8}) of 
section IIIA with $\chi > 1/2$ have an unstable mode with $k=\chi$.

\subsubsection{$k \not= \chi$}

In this subcase, the constant $k$ can be complex, and 
from Eqs.(\ref{4.5a}--\ref{4.5e}) we find that we can again integrate 
the perturbation equations yielding 
\bqn
s_{1}(z) &=& \beta  z^{1-k}
\nb\\
\sigma_{1}(z) &=& 8\pi cG \varphi_{1}(z) + \frac{1}{2}\beta(1 - k -
\chi)z^{-k}   + \sigma^{0}_{1} 
\nb\\ 
\lb{4.11a}
\varphi_{1}(z) &=& \cases{   c_{1} z^{-1/2} 
                           + c_{2} z^{-k}, & $k \neq 1/2$ \cr 
                             c_{1} z^{-1/2} 
                           + c_{2} z^{-1/2} \, \ln|x|, & $k = 1/2$ , \cr } 
\eqn
where $\beta$, $\sigma^{0}_{1}$, $c_1$, and $c_2$ are additional arbitrary 
(but now possibly complex) constants of integration.  
It is interesting to note that setting $c_{1} = 0 = s^{0}_{1}$ and $c_{2} = 
c\beta$ the above perturbations will reproduce the gauge modes given by 
Eq.(\ref{4.5gd}) that do not satisfy the boundary conditions (\ref{4.5f}) 
and (\ref{4.5g}). However, the general perturbations represented by 
Eqs.(\ref{4.11a}) will satisfy these conditions, provided that 
\bqn
\beta = \sigma^{0}_{1} & = & 0 
\nb\\  
c_{1} & = & - c_{2} \, z_{0}^{(1-2k)/2}
\nb\\
\lb{4.11b}
Re(k) & > & \frac{1}{2},\;\;\;\;\; (k \not= \chi).
\eqn

Because $Re(k) > 1/2 > 0$ for any given $\chi$, we can see that the 
solutions given by Eq.(\ref{3.8}) are unstable with an infinite 
number of unstable modes.  This result is consistent with King's 
conjecture \cite{King75}, namely: {\em Non-scalar singularities are 
unstable to perturbations and will eventually give rise to scalar 
singularities}.


\subsection{$a_{0} = 0$ and $ b_{0} \not= 0$}

In this case, the self-similar background solutions are given by
Eq.(\ref{3.16}) with $s_{0} = 1, \; \sigma^{1}_{0} = 0$. Substituting these
solutions into Eqs.(\ref{4.3a}--\ref{4.4}) we find that
\bqn
\lb{4.12a}
  z^2 s''_{1} + (2k + \chi) z s'_{1} - k(1 - k - \chi)s_{1} \qquad  & & \\
+ \, 2(k\sigma_{1} + z\sigma'_{1}) 
& = & 16\pi c G(k\varphi_{1} + z\varphi'_{1}),\\
\lb{4.12b}
z \sigma''_{1} +   (1 + k) \sigma'_{1} & = & 4\pi cG\varphi'_{1},\\
\lb{4.12c}
z s''_{1} +   \chi s'_{1} & = & 0,\\
\lb{4.12d}
z s''_{1} +   k s'_{1} & = & 0,\\ 
\lb{4.12e}
2 z\varphi''_{1} + (1+2k)\varphi'_{1}   
& = &  c s'_{1}.
\eqn
Note that in writing the above equations, we have used Eqs.(\ref{4.12c}) and 
(\ref{4.12d}) to simplify the expressions somewhat. 

Following arguments similar to those given in the previous subsection for 
ingoing solutions with $a_{0} \not= 0$ and
$b_{0} = 0$, we here impose boundary conditions on the hypersurfaces $z= z_{0}$
and $v = \infty$.  The conditions on the hypersurface $z = z_{0}$ are the same
as those given by Eq.(\ref{4.5f}).  On the hypersurface, $v = \infty \; (z =
\infty)$, we take as our condition that the perturbations should grow more
slowly than the unperturbed, 
background solution.  At future null infinity these conditions become
\bqn
\lb{4.16a}
& & s_{1}(z),\;\;   \varphi_{1}(z) \rightarrow \;\; {\rm finite},\nb\\
& & \sigma_{1}(z) \;\; {\rm grows} \;\; {\rm slower} \;\; {\rm  than}
\;\; \ln z,
\eqn
as $z \rightarrow \infty$.  As a result,
it can be shown that the gauge modes of Eqs.(\ref{4.5gb}--\ref{4.5gc}) 
are given, in the current case, by
\bq
\lb{4.16c}
\sigma_{1}(z) = c_{0} z^{-k} + c_{1},\;\;\;\;\;
 s_{1}(z) = \varphi_{1}(z) = 0.
\eq

Returning to the perturbation equations in Eqs.(\ref{4.12a}--\ref{4.12e}), we
can see from these that there are two distinct possibilities, 
$s'_{1}(z) \not = 0$ and $s'_{1}(z) = 0$.  In the former case it can be 
shown that the above equations have no solution consistent with the above 
boundary conditions at $z=z_0$ and $v\rightarrow\infty$.  Thus, in the 
following we need only consider the case $s'_{1}(z) = 0$.  In this
latter case, the perturbation equations can be rewritten as
\bqn
\lb{4.13a}
& & s_{1}(z) = s^{0}_{1},\\
\lb{4.13b}
& & z\sigma'_{1} + k \sigma_{1} = 8\pi cG(k\varphi_{1} + z\varphi'_{1})
+ \frac{1}{2}s^{0}_{1} k(1 - k - \chi),\\
\lb{4.13c}
& & 2z\varphi''_{1} + (1 + 2k)\varphi'_{1} = 0.
\eqn    
The general solution to these equations is then 
\bqn
\lb{4.15}
s_{1}(z) & = &  s^{0}_{1}
\nb\\
\sigma_{1}(z) &=&   8\pi cG \, \varphi_1(z) 
                  + \sigma^{0}_{1} \, z^{-k}
                  + \frac{s^{0}_{1}}{2} (1 - k - \chi)
\nb\\
\varphi_{1}(z) 
    &=& \cases{ \beta \ln z        + \varphi^{0}_{1},& $ k = \frac{1}{2}$,\cr
                \beta z^{(1-2k)/2} + \varphi^{0}_{1},& $ k \neq \frac{1}{2}$,\cr
              }
\eqn
where $\beta$, $\varphi_1^0$ and  $\sigma^{0}_{1}$ are integration constants
(real for $k=1/2$ and possibly complex otherwise).

Note that the gauge modes of Eq.(\ref{4.16c}) can be obtained from 
Eqs.(\ref{4.15}) by setting $\sigma_{1}^{0}=c_0$ and $s_1^0=\beta=0$.  
However, these gauge modes are 
completely eliminated by our boundary conditions (\ref{4.5f}) and 
(\ref{4.16a}). 

Applying the boundary conditions to the general 
perturbation solutions given by 
Eqs.(\ref{4.15}) eliminates all the perturbations for $k=1/2$
while for $k \not= 1/2$ there are nontrivial  
perturbations which do satisfy the boundary conditions 
provided we have  
\bqn
\lb{4.16d}
s^{0}_{1} = \sigma_{1}^{0} & = & 0 
\nb\\ 
 \varphi^{0}_{1} & = & - \beta z_{0}^{(1-2k)/2} 
\nb\\
Re(k) & > & \frac{1}{2}.
\eqn

Because the nontrivial perturbations in this case have $Re(k) > 1/2 > 0$, 
we can see 
that all the solutions in section IIIB given by Eq.(\ref{3.16}) are unstable 
with an infinite number of unstable modes.


\subsection{$a_{0}b_{0} \not= 0$}

With this most general case, we can interpret the solution as collapsing and
might hope that by examining the perturbative structure, we can determine
if there are critical solutions.  
The self-similar background solutions are given by
Eqs.(\ref{metric4}) and (\ref{eqa2}) with $\chi \ge 1/2$.  Substituting these
solutions into Eq.(\ref{4.3d})  we find that
\bq
\lb{4.17}
zs''_{1} + ks'_{1} = 0,
\eq
which has the general solution
\bq
\lb{4.18}
s_{1}(z) =\cases{\beta \ln z + s^{0}_{1}, & $ k = 1$,\cr
\beta z^{1-k} + s^{0}_{1}, & $ k \not= 1$,\cr}
\eq
where $\beta$ and $s^{0}_{1}$ are the integration constants.  

Before solving the rest of the differential equations, 
(\ref{4.3a}--\ref{4.3c}) and (\ref{4.4}), let us again first consider the boundary 
conditions. As we can see from Figs. 5 and 7, Region II of the 
spactime has only three 
boundaries: the center of the spacetime at $z = \alpha^{2}$,  the event horizon at  
$z = 0$, and past null infinity at $u = - \infty$.  In this paper we choose
$z = \alpha^{2}$ and $z = 0$ as the 
surfaces on which boundary conditions will be imposed.  

To see what kind of boundary conditions should be imposed at each of 
these surfaces, first notice that $z = \alpha^{2}$ 
is the center of the background.  It is therefore natual to 
assume that this will remain true even after the spacetime is perturbed, 
that is, $s_{1}(z = \alpha^{2}) = 0$.  In addition, because 
the background is free of spacetime singularities at the origin in 
Region II, the perturbed spacetime should remain regular there as well. 
It can be shown that this will only be the case if $\sigma_{1}(z)$ and 
the quantity $\left[(\alpha^{2} - z)\varphi'_{1} - 2k\varphi_{1}\right]$ 
remain finite.  Summarizing the above, we impose the following boundary 
conditions at the origin of coordinates  
\bqn
\lb{4.18aa}
s_{1}(y) \, \Big\vert_{y=1} & = & 0
\nb\\ 
\sigma_{1}(y) \, \Big\vert_{y=1} & \sim & \; {\rm finite}
\nb\\
\left\{(1 - y)\frac{d\varphi_{1}(y)}{d y} - 2k \varphi_{1}(y) \right\} \, 
\Bigg\vert_{y=1} & \sim & \; {\rm finite},
\eqn
where we have defined the new variable $y \equiv z/\alpha^2$ such 
that $z=\alpha^2$ corresponds to $y=1$. 

The other boundary at $z=0$, or $v = 0$, represents the event 
horizon.\footnote{Note that 
for the case $\chi \ge 1$, the hypersurface $v = 0$ is marginally trapped, as 
one can see from Eq.(\ref{3.48}).  However, in our discusssion of boundary 
conditions, we shall treat it as an event horizon too.}
We would like to be able to assume that 
no matter comes out of Region I into Region II.  Naively, we can 
express this as 
\bq\lb{4.18ab}
\phi_{,v} =   0
\eq
as $ v \rightarrow 0^{-}$.  However, it can be shown that this condition does 
not hold even for the background solutions.  Indeed, as $v\rightarrow0$, 
we have
\bq
\lb{4.18ac}
\frac{\partial \phi_{0}(u, v)}{\partial v} = 
\frac{c}{(-uv)^{1/2}\alpha(1 + y^{1/2})} \rightarrow \infty . 
\eq
But this is nothing more than the statement that in the 
coordinates $u$ and $v$, the metric coefficients are singular there. 
Having found the analytic extension of these collapsing solutions across the 
hypersurface $v = 0$ in the coodinates $\bar{u}$ and $\bar{v}$, we should 
understand the no outflow condition, Eq.(\ref{4.18ab}), in terms of 
$\bar{v}$ rather than the ``bad" coordinate $v$.   
This argument is further justified by noting that for the unperturbed, 
background solution, we have 
$\phi_{0,\bar{v}}  \rightarrow 0$, as $\bar{v} \rightarrow 0$.  In terms 
of our rescaled self-similar variable $y$, it can be shown that this 
condition is equivalent to $y^{\chi} \varphi'_{1}(y) \rightarrow 0$.  
This then, becomes our boundary condition on $\varphi(y)$ 
at $y=0$.

For the boundary conditions on the other fields at $y=0$, 
recall that when we extended the background solutions 
from Region I to Region II, we required that the extension 
be analytic.  Otherwise, the extension will not be unique and in some cases 
may even be meaningless.  To have consistency between the unperturbed and 
perturbed spacetimes, we will assume that the 
perturbations are also analytic across $y = 0$.  Given this, we find that the 
boundary conditions at the event horizon $y = 0$ become 
\bqn
\lb{4.18ad}
& & s_{1}(y),\;\ \sigma_{1}(y), \; \varphi_{1}(y) \;\; {\rm are} \; {\rm 
analytic},\nb\\
& & y^{\chi} \frac{d\varphi_{1}(y)}{dy} \rightarrow 0.
\eqn 

Again, a word about gauge modes is in order.  It can be shown that in the 
present case the gauge 
transformations (\ref{4.5ga}) lead to the perturbations 
\bqn
\lb{4.18ae}
s_{1}(y)       &=&   \beta \left(\alpha^{2}y\right)^{1-k} 
                   + s^{0}_{1}
\nb\\
\varphi_{1}(y) &=& - \frac{c \beta}{\alpha^{2k}(1 + y^{1/2})} 
                      \left(   y^{1/2 -k} 
                             - \beta^{-1}\alpha^{2(k-1)}s^{0}_{1}
                     \right) 
\nb\\
\sigma_{1}(z) &=& \frac{1}{2(1 + y^{1/2})}
                    \Bigg\{ \, \chi\left(1-y^{1/2}\right)
                                \left(   \frac{s^{0}_{1}}{\alpha^{2}} 
                                       + \frac{\beta}{\alpha^{2k}} y^{-k}
                               \right)
\nb\\
              & &            \qquad\qquad\qquad
                           + \, (1-k)\left(1 + y^{1/2}\right) 
                                \left(   \frac{s^{0}_{1}}{\alpha^{2}} 
                                       - \frac{\beta}{\alpha^{2k}} y^{-k}
                               \right)
                    \Bigg\}.
\eqn

Let us now consider the remaining  perturbation equations given 
by Eqs.(\ref{4.3a}--\ref{4.3c}) and (\ref{4.4}). It turns out to be 
convenient to consider the two cases $k =1$ and $k \not= 1$ separately.

\subsubsection{$k = 1$}

In this case from Eq.(\ref{4.18}) we find that 
\bq
\lb{4.18a}
s_{1}(y) = \beta \ln\left(y\right).
\eq
where, in writing the above expression, we have chosen 
$s^{0}_{1}  = -2\beta\ln(\alpha)$ so that the boundary condition $s_{1}(y = 
1) = 0$ is satisfied.  Inserting Eq.(\ref{4.18a}) into 
Eqs.(\ref{4.3a}--\ref{4.3c}) and (\ref{4.4}), 
we find that only two of the four equations are independent  
\bqn
\lb{4.19a}
\sigma_{1}(y) &=&   8\pi c G\left(1 - y^{1/2}\right)
                         \left[   y^{1/2}\left(1 + y^{1/2}\right)
                                          \frac{d\varphi_{1}(y)}{dy} 
                                + \varphi_{1}
                        \right]
\nb\\ 
              & & - \frac{\beta}{2\alpha^{2}y(1 + y^{1/2})^{2}}
                         \Bigg[    \, (1-\chi)(1 + y^{2}) 
\nb\\
              & &                  \qquad\qquad\qquad\qquad
                                + \, 2(1+\chi)y 
                                +    2 y^{1/2}(1 + y)
                         \Bigg] 
   \\
\lb{4.19b}
 y(1-y)\frac{d^{2}\varphi_{1}}{dy^{2}} 
&+& \left[\lambda - (a+b+1)y\right]\frac{d\varphi_{1}}{dy}  
- ab \varphi_{1}  = f(y),
\eqn
where 
\bqn
\lb{4.19ba}
a &=& 1, \;\;\; b = \frac{1}{2},\;\;\; \lambda = \frac{3}{2},\nb\\
f(y) &\equiv& \frac{c\beta}{2\alpha^{2}y(1 + y^{1/2})^{2}}
 \left[(1-y) - y^{1/2}\ln(y)\right].
\eqn
Eq.(\ref{4.19b}) is the inhomogeneous hypergeometric equation \cite{Erd}, and 
the corresponding general solution is given by
\bq
\lb{4.20}
\varphi_{1}(y) = \varphi_{1}^{h}(y) + \varphi_{1}^{s}(y),  
\eq
where $\varphi_{1}^{s}(y)$ is a particular solution of the inhomogeneous 
equation (\ref{4.19b}), and $\varphi_{1}^{h}(y)$ is the general solution of 
the 
associated homogeneous equation,
\bq
\lb{4.20a}
y(1-y)\frac{d^{2}\varphi_{1}}{dy^{2}} 
+ \left[\lambda - (a+b+1)y\right]\frac{d\varphi_{1}}{dy}  
- ab \varphi_{1} = 0.
\eq
The general solution $\varphi^{h}_{1}(y)$ can be written 
as a linear combination of two independent solutions $\varphi^{1}_{1}$ and 
$\varphi^{2}_{1}$ 
\bq
\lb{4.20c}
\varphi_{1}^{h}(y) = c_{1} \, \varphi^{1}_{1}(y) + c_{2} \, \varphi^{2}_{1}(y), 
\eq
where $c_{1}$ and $c_{2}$ are two arbitrary constants, and $\varphi^{1}_{1}$ 
and $\varphi^{2}_{1}$ are given 
by
\bqn
\lb{4.20d}
\varphi^{1}_{1} &=& 2 F(1,\frac{1}{2};\frac{3}{2}; y)= 
\frac{1}{y^{1/2}}\ln\left(\frac{1 + y^{1/2}}{1 - y^{1/2}}\right),
\nb\\
\varphi^{2}_{1} &=& y^{-1/2}F(\frac{1}{2}, 0; \frac{1}{2};y) = y^{-1/2},
\eqn
with $F(a,b;\lambda;y)$ denoting the hypergeometric function.
A particular solution of the inhomogeneous equation (\ref{4.19b}) can
be found and is given by
\bqn
\lb{4.20e}
\varphi^{s}_{1}(y) & = & \frac{c\beta}{\alpha^{2}(1+y^{1/2})}
                              \left[   \ln y  
                                     - 2\left(\frac{1+ y^{1/2}}{y^{1/2}}\right)
                                        \ln \left(\frac{1+y^{1/2}}{2}\right)
                             \right].
\eqn 
From Eqs.(\ref{4.19ba}--\ref{4.20e}), we can show that as $y \rightarrow 1$,
we have 
\bq
\lb{4.20f}
\varphi_{1}(y) \approx   \frac{1}{y^{1/2}}\left[c_{1} 
\ln\left(\frac{1 + y^{1/2}}{1 - y^{1/2}}\right) 
+ c_{2}\right].
\eq
As a result, the boundary condition for 
$\varphi_{1}$ in Eq.(\ref{4.18aa}) holds only 
if $c_{1} = 0$ while the condition for $\sigma_{1}(y)$ at $y=1$ 
does not impose any additional restrictions. 

With regard to the boundary conditions at $y=0$, we find from Eqs.(\ref{4.18a})
and (\ref{4.20e}) that the 
conditions on $s_{1}(y)$ and $\sigma_{1}(y)$ further demand $c_{2} = \beta = 
0$.  Therefore, in the case that $k=1$, the boundary conditions 
(\ref{4.18aa}) and (\ref{4.18ad}) eliminate all the perturbations.

\subsubsection{$k \not= 1$}

When $k \not= 1$, we have
\bq
\lb{4.21}
s_{1}(y) = c_{0} \alpha^{2} \left(y^{1-k} - 1\right),  
\eq
where $c_{0} \equiv \alpha^{-2k}\beta$. Note that now the constant $k$ can be 
complex, as can $c_{0}$. Substituting 
Eq.(\ref{4.21}) into Eqs.(\ref{4.3a}--\ref{4.3c}) and (\ref{4.4}), we find 
that now there are also only two independent equations, which can be written 
as
\bqn 
\lb{4.22a}
k \, \sigma_{1}(y) &=&   \frac{\chi}{c}(1 - y^{1/2}) 
                            \left[   y^{1/2}(1 + y^{1/2})
                                        \frac{d\varphi_{1}(y)}{dy} 
                                   + k \varphi_{1}
                           \right]
\nb\\
                   & & - \,\frac{c_{0}(1-k)}{2(1 + y^{1/2})}
                             \bigg[ \, \left(k + \chi\right)
                                       \left(1 + y^{1/2 -k}\right)
\nb\\
                   & &                 \qquad\qquad\qquad\qquad
                                    +  \left(k - \chi\right)
                                       \left(y^{1/2} + y^{-k}\right)
                             \bigg]
   \\
\lb{4.22b}
 y(1 - y)\frac{d^{2}\varphi_{1}}{dy^{2}} 
& + & \left[\lambda - (a+b+1)y\right]\frac{d\varphi_{1}}{dy}
- ab\varphi_{1} = f(y),
\eqn
but now with
\bqn
\lb{4.22ba}
a &=& k,\;\;\;\;
b = \frac{1}{2},\;\;\;\;
\lambda = k + \frac{1}{2},\nb\\
f(y) &\equiv& \frac{c\beta}{2\alpha^{2k}y^{k + 1/2}(1 + y^{1/2})^{2}}
\left[(1-k)(1 - y^{k})y^{1/2} - ky(1 - y^{k-1})\right].
\eqn
Similar to the previous case, the general solution of Eq.(\ref{4.22b}) can be 
written in the same forms as those given by  Eqs.(\ref{4.20}--\ref{4.20c}), 
but $\varphi^{1}_{1}$ and $\varphi^{2}_{1}$ now are given by $\varphi^{1}_{1}=
F(\frac{1}{2}, k; k + \frac{1}{2}; y)$, and $\varphi^{2}_{1} = F(\frac{1}{2}, 
k; 1; 1-y)$.  In addition, it can be shown that a particular solution of the 
inhomogenuous equation (\ref{4.22b}) is given by
\bq
\lb{4.25}
\varphi^{s}_{1}(y) = - \frac{c\beta}{\alpha^{2k}} \, 
                       \frac{1+ y^{1/2 - k}}{1 + y^{1/2}} . 
\eq
Thus, the general solution for $\varphi_{1}(y)$ in the 
present case is given by
\bqn
\lb{4.26}
\varphi_{1} (y) &=&   c_{1} \, F(\frac{1}{2}, k; k + \frac{1}{2}; y) 
                    + c_{2} \, F(\frac{1}{2}, k; 1; 1-y) 
                    - \frac{c\beta}{\alpha^{2k}} \, 
                      \frac{1+ y^{1/2 - k}}{1 + y^{1/2}} .                
\eqn
From Eqs.(\ref{4.21}), (\ref{4.22a}) and (\ref{4.26}) we can see that when 
$c_{1} = c_{2} = 0$, the perturbations reduce to those given in
Eq.(\ref{4.18ae}) and which are purely gauge modes. Therefore, in the 
following we will restrict ourselves to those solutions for which 
$\left|c_{1}\right|^{2} + \left|c_{2}\right|^{2} \not= 0$.

Considering first the boundary conditions at the origin, $y = 1$, we note
the following limiting relations as $y \rightarrow 1$  
\bqn
\lb{4.27}
F(\frac{1}{2}, k; k + \frac{1}{2}; y) 
   & \rightarrow & - \frac{1}{2}(2k-1)A(k)\ln(1-y) + A_{1}(y;k) 
\nb\\
F(\frac{3}{2}, k +1; k + \frac{3}{2}; y) 
   & \rightarrow &   \frac{1}{2}(4k^{2} -1)A(k) \left(   \frac{2}{k(1-y)} 
                                                       + \ln(1-y)
                                                \right) 
                   + A_{2}(y;k) 
\eqn
where $A(k) \equiv \Gamma(k - 
\frac{1}{2})/\left(\Gamma(k)\Gamma(\frac{1}{2})\right)$, and $A_{1}(y;k)$ and 
$A_{2}(y;k)$ are the finite pieces of the hypergeometric functions in the 
limit as $y \rightarrow 1$. Then, from Eqs.(\ref{4.26}--\ref{4.27}) and 
the relation
\bq
\lb{4.28}
\frac{d}{dz}F(a, b; \lambda; z) = \frac{ab}{\lambda}F(a+1, b+1; \lambda+1; z),
\eq
we find that the boundary conditions given by Eq.(\ref{4.18aa}) at $y=1$ 
will hold, provided that
\bq
\lb{4.29}
c_{1} = 0.
\eq

For the boundary conditions at $y=0$ as given in Eq.(\ref{4.18ad}), we 
first recall that our imposition of analyticity on the extended spacetime
exploited the barred coordinate system.  Doin the same here, we notice that 
$y = \alpha^{-2}v/u = \alpha^{-2}(\bar{v}/\bar{u})^{2n}$ and 
\bq
\lb{4.30}
s_{1}(y) = c_{0}\alpha^{2}\left(\alpha^{-2}\left(\frac{\bar{v}}
{ \bar{u}}\right)^{2n(1-k)} - 1\right).
\eq
Thus, to have $s_{1}(y)$ be analytic across the boundary $y=0$ ({\it i.e.} 
$\bar{v} = 0$), we must have
\bq
\lb{4.31}
k = \frac{m}{2n} < 1,
\eq
where $m$ is a positive integer.\footnote{In principle, $m$ could be a negative
integer as well, but our interest here is in unstable modes with $k>0$. 
Hence we restrict to $m>0$.}
Due to the properties of the hypergeometric functions, it is convenient
to consider separately the cases $k = l + \frac{1}{2}$ and $ k 
\not= l + \frac{1}{2}$, where $l$ is a non-negative integer.

\bigskip
\noindent {\bf Case (a) $\; k = l + \frac{1}{2}$}: In this case, as 
$y \rightarrow 0$, we find that
\bqn
\lb{4.31a}
F(\frac{1}{2}, k; 1; 1-y) 
  & \rightarrow & \frac{ (l-1)! \, y^{-l} }
                       { \Gamma(\frac{1}{2} + l)\Gamma(\frac{1}{2})} 
                  \, \sum^{l-1}_{p=0}{\frac{(a-l)_{p}(b-l)_{p}}
                                           {(1-l)_{p} \, p!} \, y^{p}}
\nb\\
& & - \frac{(-1)^{l}}{\Gamma(\frac{1}{2} - l)\Gamma(\frac{1}{2})} \, \ln y.
\eqn
Inserting the above expression into Eq.(\ref{4.26}) and 
considering Eq.(\ref{4.29}), one quickly sees that the condition 
that $\varphi_{1}(y)$ 
be analytic across the hypersurface $y=0$ can only be true for  
$k < 1/2$ and $c_{2} = 0$.
Because the boundary conditions as $y=0$ force us to take $c_{1} = c_{2} = 0$, 
we can see that the corresponding 
perturbations are purely gauge.  Thus, we need no longer consider the 
case of $k=l+\frac{1}{2}$.

\bigskip

\noindent {\bf Case (b) $\; k \not= l + \frac{1}{2}$}: In this case, 
it can be shown that in the limit $y \rightarrow 0$ 
\bqn
\lb{4.32}
F(\frac{3}{2}, k + 1; 2; 1- y)  
  & \rightarrow &   k^{-1}A(k) \bigg[   (2k -1)y^{-1/2 -k}  
                                      - (1-k) y^{1/2 - k}
                               \bigg] 
                  + A_{3}(y;k) 
\eqn
where $A_{3}(y;k)$ is again the finite portion of the hypergeometric portion 
in the limit as $y \rightarrow 0$.  Inserting this into the general solution 
for $\varphi_1(y)$ as given in Eq.(\ref{4.26}) together with the fact that 
we still have $c_{1} = 0$, we can find the limiting behavior of the derivative
of $phi_1(y)$ as $y \rightarrow 0$, namely
\bqn
\lb{4.33}
\frac{d\varphi_{1}(y)}{dy} 
  & \rightarrow &   \frac{1}{2} \Big\{   \, cc_{0}\left[   y^{-1/2}
                                                         + 2(1-k)y^{-k} \right]
\nb\\
  &  &                \qquad\qquad       + c_{2}(2k-1)A(k) y^{-1/2 - k}
\nb\\
  &  &                \qquad\qquad       + \left[   cc_{0}(10k -7) 
                                                  - c_{2}(1-k)A(k)
                                          \right] \, y^{1/2-k}
                                \Big\} 
\nb\\
  &  &                \qquad\qquad       
                  + A_{4}(k) 
\eqn
where, again, $A_{4}(y;k)$ is the finite part.  Now, one 
can see that the boundary condition of vanishing of $y^{\chi} \, \varphi'_{1}(y)$
at $y=0$ can only be true if 
\bq
\lb{4.34}
k < \chi - \frac{1}{2} \;\; {\rm and} \;\;
\chi > \frac{1}{2} 
\eq
because we want to demand that $c_{2} \not= 0$.  Otherwise, the perturbations
would again be purely gauge. 

We are not yet done as we must still consider the effect of imposing 
analyticity on $\sigma_1(y)$ at $y=0$.  From Eq.(\ref{4.32}) and the 
asymptotic behavior of 
\bq
\lb{4.35}
F(\frac{1}{2}, k; 1; 1- y) \rightarrow A(k) y^{1/2 - k} + A_{5}(k),
\eq
as $y \rightarrow 0$, we get that
\bqn
\lb{4.36}
\sigma_{1}(y)  
   & \rightarrow &   \frac{\chi}{c} \left( A(k)c_{2} - c c_{0} \right) \, 
                       y^{1/2 - k}  
                   + \frac{1}{2} \Big[   A(k)\chi(2k-1)c_{2} 
                                       - (1-\chi -k)c_{0}
                                 \Big] \, y^{-k} 
                   + \sigma_{1}^{0} 
\eqn
as $y \rightarrow 0$.  Thus, to have $\sigma_{1}(y)$ be analytic across 
the hypersurface $y = 0$, we must assume
\bq
\lb{4.37}
k < \frac{1}{2} \;\; {\rm and} \;\;
 c_{2} = \frac{1-\chi -k}{\chi(2k-1)A(k)}c_{0}.
\eq
When $k = 1 - \chi$ ({\it i.e.} $m = 1$), we can see from the above expression 
that $c_{2} = 0$.  In other words, the mode $m = 1$ is a pure gauge mode.  
Thus, physically relevant perturbations will have $m \not= 1$. 

Combining Eqs. (\ref{3.39}), (\ref{4.31}) and (\ref{4.34}) with 
Eq.(\ref{4.37}),   we find that in the present case all the perturbations with
\bqn
\lb{4.38}
& & c_{1} = 0,\;\;\; c_{2} = \frac{n(1-m)\Gamma(k)\Gamma(\frac{1}{2})}
{(m-n)(2n -1)\Gamma(k - \frac{1}{2})}c_{0},\nb\\
& & k = \frac{m}{2n},\;\;\; 1 < m < n -1,\;\;\;\;\; n > 1,
\eqn
satisfy the boundary conditions (\ref{4.18aa}) and (\ref{4.18ad}). 

Therefore, {\em for any parameter $n>2$, the corresponding solution of 
Eqs.(\ref{metric4}) and (\ref{eqa2})  has $(n-3)$ 
unstable modes}. In particular, the one with $n = 4$  only has one unstable 
mode. By definition this solution represents a critical solution \cite{Gun00}. 
This is consistent with Garfinkle's original observations \cite{Gar00}.
However, it is a departure from the perturbation analysis of \cite{GG02} 
where they calculate the $n=2$ solution to have a single unstable mode.  
The discrepancy between that work and ours would seem to lie in the choice
of boundary conditions for the perturbation problem.  Though both works 
consider the same boundaries at the origin ($y=1$) and the horizon ($y=0$), 
we have the added condition that nothing escape from the horizon.  This
would seem to eliminate some of the additional modes found in \cite{GG02}.


\section{ Critical Collapse of Scalar Field With Potential }

In this section, following \cite{GMG96} we argue that the 
critical solution with $n = 4$ (or $\chi = 7/8$) for the massless scalar field 
collapse found in the last section is also the critical solution for the collapse of a 
scalar field with any potential $V(\phi)$ and  cosmological 
constant $\Lambda$, provided that the condition (\ref{3.4}) holds. As shown in 
section III, because of the presence of these terms, globally self-similar 
solutions do not exist, although in the limit $\tau 
\rightarrow \infty$ asymptotically self-similar solutions do exist. Fortunately 
for the study of critical collapse,  
this is sufficient. Indeed, initial configurations for collapsing 
gravitational systems  
can be quite different. This is the case even in situations where globally self-similar 
solutions are permitted. Thus, in generic cases the spacetime will become   
self-similar only after the passage of a sufficent amount of time 
during which the original  
differences will be washed away and the collapse approaches a 
self-similar evolution, such that the physics on large scales is related to that on small 
scales in a universal manner.  A schematic illustration of this process is 
given in Fig. 8.  Region $A$
represents a region where the influence of the terms $V(\phi)$ and $\Lambda$ 
in Einstein field equations is still very strong, and the corresponding 
spacetime is significantly  
different from that of the critical solution.  In particular, no trapped 
surfaces have been formed so far. As $\tau$ grows, the differences are 
smoothed out, and when $\tau$ grows to a certain value, say, $\tau = 
\tau_{1}$, the differences become very small and can be considered as perturbations on top of the 
the self-similar critical solution. The region where such perturbations are 
valid is denoted as region $B$ in Fig. 8. Clearly, the time $\tau_{1}$  is 
dependent on the initial-data, and by fine tuning one can find out which initial data 
leads to critical collapse and which does not.  When the strength of the 
initial data is larger than that required for critical collapse, the collapse
will be strong enough to form black hole, while when weaker, the scalar field 
will reflect through the origin before a black hole can be formed and 
subsequently disperse entirely to infinity.  In region 
$B$, the solutions to the whole problem can be expanded in the form 
\cite{GMG96},
\bq
\lb{5.1}
A(\tau, z) = \sum^{\infty}_{n = 0}{e^{- n\tau} A^{(n)}_{*}(z)},
\eq
where $A = (\sigma, \; s,\; \varphi)$, and $A^{(0)}_{*}$ denotes the critical 
solution.
Substituting the above expression into the Einstein field equations 
(\ref{2.13a}--\ref{2.13e}), and then separating terms by powers of 
$e^{-\tau}$, we find that, to zeroth order, the resultant equations are 
exactly those given by Eqs.(\ref{3.5a}--\ref{3.5e}) for the self-similar 
critical solution $A^{(0)}_{*}$, which are independent of other higher terms. 
On the other hand, the equations for $A^{(n)}_{*}$'s with $n \ge 1$ are 
always coupled to its lower terms, $A^{(n-1)}_{*}, \cdots, A^{(0)}_{*}$. 
This allows us 
to determine $A^{(n)}_{*}$ recursively starting with $A^{(0)}_{*}$. Thus, in 
principle once $A^{(0)}_{*}$ is known, we can find all the higher terms 
$A^{(n)}_{*}$. In particular, one can show that $A^{(1)}_{*}$ satisfies the 
same equations as those for $A_{1}(z)$ given by Eq.(\ref{4.3a}--\ref{4.4}) 
with $k = -1$, where $A_{1}(z) = (s_{1},\; \sigma_{1},\; \varphi_{1})$. 
Following the analysis given there, we find that $A^{(1)}_{*}$ is well-behaved 
in region $B$, regular at $r = 0$, and analytical across the hypersurface 
${v} = 0$. Since the higher order terms $A^{(n)}_{*}$ satisfy similar 
types of differential equations as those for $A^{(1)}_{*}$, and also these 
higher terms in (\ref{5.1}) all correspond to  stable modes  ($k < 0$) of 
perturbations given by Eq.(\ref{4.1}), it is quite reasonable to expect that 
all the higher order terms in Eq.(\ref{5.1}) are well behaved in Region $B$, 
and in the limit $\tau \rightarrow \infty$ Eq.(\ref{5.1}) converges to  
$A^{(0)}_{*}$. Note that $A^{(0)}_{*}$ is well-behaved in the whole region 
$B$, except on the hypersurface ${v} = 0$. However, this divergence is weak 
(logarithmically) and is due entirely to the choice of coordinates. As shown in 
section III, in terms of $\bar{u}$ and $\bar{v}$ the metric cofficients are 
well behaved there. In review of all the above, we can see that Eq.(\ref{5.1}) 
indeed represents solutions to the full problem, that is, solutions of 
Einstein-scalar equations with potential $V(\phi)$ and the cosmological 
constant $\Lambda$.

\begin{figure}[htbp]
\begin{center}
\label{fig8}
\leavevmode
    \epsfig{file=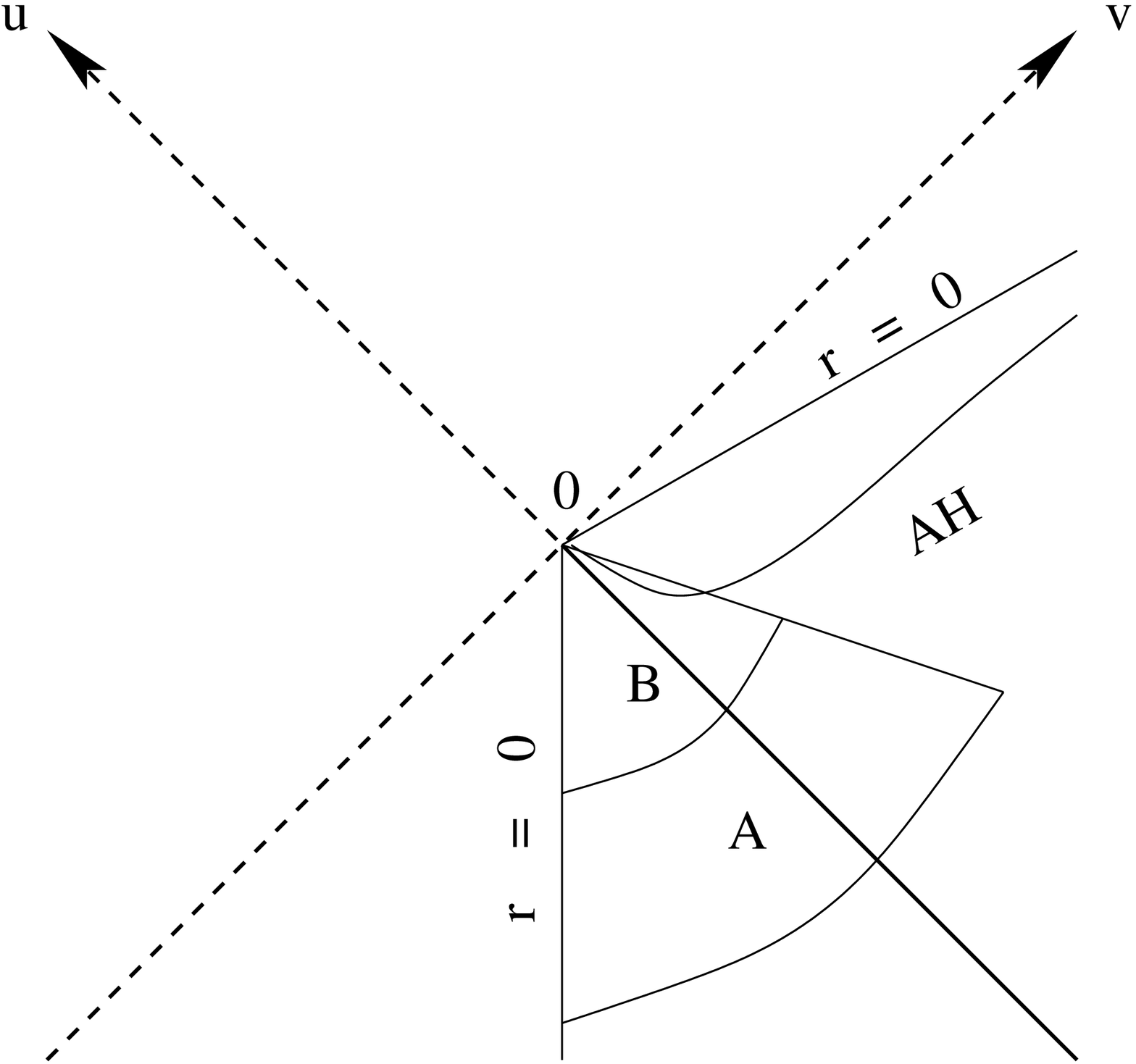,width=0.5\textwidth,angle=0}
\caption{Global structure of near-critical gravitational collapse. In region $A$ the effects 
of the terms $V(\phi)$ and $\Lambda$ are strong and the spacetime will, in 
general be very 
different from that of the critical solution. In particular, an apparent 
horizon will not have formed.  With increasing time, however, the effects 
of the potential will become more and more weak such that,  
in region $B$ and for those initial configurations with initial data close to 
criticality, those effects can be considered as perturbations of the 
critical solution.  In this region, supercritical configurations 
will form, in the end, black 
holes, while subcritical configurations will reflect through the origin 
and disperse to infinity without formating black holes. The line $AH$ 
represents  the apparent horizon formed in the supercritical collapse.  }      
\end{center} 
     
\end{figure}

Now let us consider the linear perturbations of the solutions (\ref{5.1}),
which can be written as
\bq
\lb{5.2}
\delta A(\tau, z) = e^{k\tau}\delta_{k} A(\tau, z),
\eq
where $\delta_{k} A(\tau, z)$ denotes the general perturbations of $A(\tau, 
z)$ for the mode $k$, similar to that given by Eq.(\ref{4.1}), but now 
depending on both of $\tau$ and $z$. However, we can expand it in   a similar 
form as that given by Eq.(\ref{5.1}),
\bq
\lb{5.3}
\delta_{k} A(\tau, z) = \sum^{\infty}_{n = 0}{e^{- n\tau} 
\delta_{k}A^{(n)}(z)}.
\eq
As $A(\tau,z)$ is  completely determined by $A^{(0)}_{*}$, $\delta_{k} A(\tau, 
z)$ is totally specified by its leading term $\delta_{k}A^{(0)}(z)$.
Indeed, substituting Eqs.(\ref{5.1}--\ref{5.3}) into the Einstein field 
equations 
(\ref{2.13a}--\ref{2.13e}), we find that $\delta_{k}A^{(0)}(z)$ satisfies
the same differential equations (\ref{4.3a}--\ref{4.4}) for $A_{1}(z)$, where 
the background depedence is only on terms of $A^{(0)}_{*}$. Consequently, the 
spectrum for the perturbations of $\delta_{k}A^{(0)}(z)$ is the same as that 
of $A_{1}(z)$.  Once $\delta_{k}A^{(0)}(z)$ is known we can find the higher 
order terms $\delta_{k}A^{(n)}(z)$ recursively. In this way, we can see that 
the spectras of perturbations in the cases with or without $V(\phi)$ and 
$\Lambda$ will be exactly the same. In particular,   {\em the critical 
solution for the massless scalar field   is also a critical solution for   a 
scalar field with potential in a non-vanishing cosmological constant 
background (with the possibilities of  being either negative or positive),   
provided Eq.(\ref{3.4}) holds}.

\section{Conclusions}

In this paper, we have considered the Einstein field equations for a 
scalar field with potential $V(\phi)$ in a background where the 
cosmological constant $\Lambda$ is different from zero.  We have argued 
that $V(\phi)$ and $\Lambda$ will not affect the critical behavior 
provided the condition (\ref{3.4}) is true.

All the continuously self-similar solutions to Einstein-massless-scalar field 
equations were found together with their analytic extensions.  It was shown
that some of these solutions represent gravitational collapse of the scalar 
field.  In all these collapsing models, a black hole is always formed.  It 
should be noted that recently Ida showed that no black holes can be formed 
in (2+1) dimensional gravity, unless the dominant energy condition is 
violated \cite{Ida00}.  This seems to conflict with the results presented in 
this paper. One possible explanation may be that the conditions assumed in 
\cite{Ida00} do not hold here, specifically the one that the event horizon is a 
compact surface. 

In section IV, we also studied the linear perturbations of all the 
self-similar solutions discussed in section III, and found all the 
perturbations in closed form. 
On imposing boundary conditions of (1) regularity at the 
origin and (2) analyticity and an inflow only condition at the event horizon,
we were able to identify the unique, exact solution which has a 
single unstable mode.  This solution is that given by Eqs.(\ref{metric4}) 
and (\ref{eqa2}) with $n = 4$, and which was first found by 
Garfinkle \cite{Gar00}, and which he showed matches well with the 
critical solution found numerically by Pretorius and Choptuik \cite{PC00}. 
While this result is certainly very encouraging, we must note that in 
our case the exponent $\gamma$ that can be obtained from the expression 
$\gamma = 1/k = 4$, is significantly different from the one determined 
numerically by Pretorius and Choptuik, $\gamma \sim 1.2 \pm 0.05$.  Nor is
it similar to the critical exponent of $\gamma \sim 0.81$ which 
Husain and Olivier \cite{HO01} find using a different 
numerical code written in double null coordinates.  
The recent perturbative results of Garfinkle and Gundlach \cite{GG02}, 
while finding the $n=2$ solution to be the critical solution, do get 
a better critical exponent, namely $\gamma = 4/3 \approx 1.33$.  All of these 
differences seem quite intriguing and together with the problem as a whole, are 
worthy of further investigation.

\section*{Acknowledgmentes}  

The authors (AW $\&$ YW) would like to thank the Department of Physics and 
Astronomy and the Department of Mathematics at BYU for hospitality while 
this work was completed.  They would also like to  
acknowledge the financial support of NSF Grant PHY-9900644 (EWH) to 
BYU, the support of CNPq-Brazil and FAPERJ-Brazil (AW), and a postdoctoral 
fellowship from CNPq-Brazil (YW).


\section*{Appendix: Exact Solutions for Massless Scalar Field}

\renewcommand{\theequation}{A.\arabic{equation}}
\setcounter{equation}{0}

As shown in section II, when $\tau \rightarrow \infty$, the
terms associated with the potential $\tilde{V}(\phi)$ always go like 
$e^{-2\tau}\tilde{V}(\phi)$.
Thus, for the cases where $\tilde{V}(\phi)$ is finite, the potential terms
are negligible in the strong field regime, where critical behavior becomes
visible. Setting $\tilde{V}(\phi) = 0$ in Eqs.(\ref{2.7}--\ref{2.8}), we find
that the corresponding Einstein-scalar  field equations  become, 
\bqn 
\lb{A.1a}
 r_{,uu} - 2\sigma_{,u} r_{,u} &=& - 8\pi G r \phi_{,u}^{2},\\
\lb{A.1b}
 r_{,vv} - 2\sigma_{,v} r_{,v} &=& - 8\pi G r \phi_{,v}^{2},\\
\lb{A.1c}
r_{,uv} + 2r\sigma_{,uv}  &=& - 8\pi G r  \phi_{,u}\phi_{,v},\\
\lb{A.1d}
r_{,uv}    &=&  0,\\
\lb{A.1e}
 2\phi_{,uv} + \frac{1}{r}\left(r_{,u}\phi_{,v} + r_{,v}\phi_{,u}\right)
&=& 0.
\eqn
From Eq.(\ref{A.1d}) we find that the metric coefficient $r$  has the 
general solution, 
\bq
\lb{A.2}
r = \alpha(u) + \beta(v),
\eq
where $\alpha(u)$ and $\beta(v)$ are arbitrary functions of their indicated
arguments. Clearly, in general, the solutions can be divided into three
different cases, 
\bq
\lb{A.2a}
(i) \; \alpha'(u) = 0, \;\; \beta'(v) \not= 0;\;\; 
(ii) \; \alpha'(u)  \beta'(v) \not= 0;\;\;
(iii) \; \alpha'(u) \not= 0, \;\; \beta'(v) = 0.
\eq
  In the cases $(i)$ and $(iii)$, by exchanging the two
variables $u$ and $v$, we can get one from the other. Therefore, without loss
of generality, in the following we shall  consider only Cases $(i)$ and
$(ii)$.

\subsection*{A.1 Exact Solutions for $\alpha'(u) = 0,\;\beta'(v) \not=0$}

In this case we have $\alpha(u) = $ const. By redefining the function
$\beta(v)$, we can always set this constant to zero. Then, from
Eqs.(\ref{A.1a}) and (\ref{A.1c}) we find that 
\bq
\lb{A.3}
\phi = \phi(v),\;\;\;
\sigma = a(u) + b(v),
\eq
where $a(u)$ and $b(v)$ are other two arbitrary functions. 
However, by introducing  new coordinates $\bar{u}$ and $\bar{v}$ via the
relations, 
\bq
\lb{A.4}
d\bar{u} = e^{2a(u)}du, \;\;\;\;
d\bar{v} = e^{2b(v)}dv,
\eq
we can always set $a(u) = 0 = b(v)$, a choice that will be assumed in
the following discussion. Then, substituting   Eq.(\ref{A.3})   into
Eq.(\ref{A.1b}),  we find that   
\bq 
\lb{A.5} 
\phi(v) = \pm \frac{1}{\sqrt{8\pi G}}\int{\left[-
\frac{\beta''(v)}{\beta(v)}\right]^{1/2} dv} + \phi^{1}_{0}.
\eq
Thus, for any given function  $\beta(v)$, Eq.(\ref{A.5})
 completely determines the massless scalar field, provided that    
\bq 
\lb{A.6}
 \beta(v) \ge 0,\;\;\;\; \beta''(v)\le 0.
\eq
On the other hand, in order not to have the spacetime be closed at the
beginning of collapse, we also require that
\bq
\lb{A.7}
\beta'(v) > 0, \; (v > v_{0}),
\eq
for some value of $v_{0}$, where $v_{0}$ is a constant.
Combining Eqs.(\ref{A.6}) and (\ref{A.7}), it can be shown
that in the present case the collapse {\em never} forms black holes. This is
because the formation of black hole is indicated by the formation of an
apparent horizon, on which we have  
\bq
\lb{A.8}
\beta'(v_{AH}) = 0.
\eq
However,   Eqs.(\ref{A.7}) and (\ref{A.8}) with the condition
that $\beta(v) \ge 0$ imply that $\beta''(v) > 0$ right outside the apparent
horizon. The latter is possible only when $\phi$ is imaginary, which is
forbidden by the weak energy condition \cite{HE73}. 

It should be noted that, although the collapse cannot form black holes, it can
form singularities that are null and strong. To show this, let us first notice
that in the present case  all the scalars built from the Riemann curvature
tensor are zero, therefore, scalar curvature singularities are always absent 
\cite{ES77}.  However, we do have non-scalar curvature singularities at $r =
0\;$ (or equivalently  $\beta(v) = 0$), in the sense that the tidal forces
experienced by an observer become infinitely large as the hypersurface $r = 0$
approaches. To this end, let us consider the radial timelike geodesics,
which in the present case are simply given by
\bq
\lb{A.9}
\dot{u} = e^{-2d_{0}},\;\;\;\;
\dot{v} = \frac{1}{2}e^{2d_{0}},\;\;\;\;
\dot{\theta} = 0, 
\eq
where $d_{0}$ is an integration  constant, and an overdot denotes the
ordinary derivative with respect to the proper time, $\lambda$, of the timelike
geodesics. Defining $e^{a}_{(0)} = d{x}^{a}/d\lambda$, we find that the three
unit vectors, given by
\bqn
\lb{A.10}
e^{a}_{(0)} &=& e^{-2d_{0}}\delta^{a}_{u} + 
\frac{1}{2}e^{2d_{0}}\delta^{a}_{v},\nb\\
e^{a}_{(1)} &=& e^{-2d_{0}}\delta^{a}_{u} - 
\frac{1}{2}e^{2d_{0}}\delta^{a}_{v},\nb\\
e^{a}_{(2)} &=& \frac{1}{r} \delta^{a}_{\theta},
\eqn
form a freely falling frame,  
\bq
\lb{A.11}
e^{a}_{(c)}e^{b}_{(d)} g_{ab} = \eta_{cd},\;\;\;
e^{a}_{(c); b}e^{b}_{(0)}  = 0,
\eq
where $\eta_{ab} = {\mbox{diag.}}\;\{-1, \;1,\; 1\}$. 
Projecting the Ricci tensor onto the above frame, we find that
\bqn
\lb{A.12}
R_{(a)(b)} = R_{cd}e^{c}_{(a)}e^{d}_{(b)} &=& -
\frac{1}{4}e^{4d_{0}}\left[\frac{\beta''(v)}{\beta(v)}\right]\nb\\
& & \times \left[\delta^{0}_{a}\delta^{0}_{b} -
\left(\delta^{0}_{a}\delta^{1}_{b} + \delta^{1}_{a}\delta^{0}_{b}\right)
+ \delta^{1}_{a}\delta^{1}_{b}\right].
\eqn
To study the asymptotic behavior of the solutions near the
hypersurface $r = \beta(v) = 0$, we expand the function $\beta(v)$ near
this surface and assume that  the leading term of the expansion is
proportional to $v^{\gamma}$, i.e., 
\bq
\lb{A.13} \beta(v) = \beta_{0} v^{\gamma}, 
\eq
where the constants $\beta_{0}$ and $\gamma$ have to satisfy the  conditions
$\beta_{0} > 0$ and $0 < \gamma < 1$, in order to have  
Eqs.(\ref{A.6}) and (\ref{A.7}) hold. Substituting Eq.(\ref{A.13}) into
Eq.(\ref{A.12}), we find that
\bqn
\lb{A.14}
R_{(a)(b)} &=&
\frac{1}{4v^{2}}\gamma(1-\gamma)e^{4d_{0}} \left[\delta^{0}_{a}\delta^{0}_{b} -
\left(\delta^{0}_{a}\delta^{1}_{b} + \delta^{1}_{a}\delta^{0}_{b}\right) +
\delta^{1}_{a}\delta^{1}_{b}\right]\nb\\
&=& \frac{\gamma(1-\gamma)}{\lambda^{2}} 
\left[\delta^{0}_{a}\delta^{0}_{b} - \left(\delta^{0}_{a}\delta^{1}_{b} +
\delta^{1}_{a}\delta^{0}_{b}\right) + \delta^{1}_{a}\delta^{1}_{b}\right],
\eqn
where  the proper time $\lambda$ along the timelike geodesics was chosen such
that $r = 0$ corresponds to $\lambda = 0$. The above equation shows clearly
that $R_{(a)(b)}$ becomes unbounded as $r \rightarrow 0$. Thus, the
hypersurface $r = 0$ represents a non-scalar spacetime singularity. This
singularity is   strong \cite{Newmann96}, in the sense that the distortion,
which is proportional to two integrals of $R_{(a)(b)}$ with respect to
the proper time $\lambda$, becomes unbounded 
\bq
\lb{A.15}
\int{d\lambda\int{R_{(a)(b)} d\lambda}} \; \sim \; \gamma(\gamma - 1)\ln
(\lambda) \; \rightarrow \; \infty,
\eq
as $\lambda \rightarrow 0$. Note that this is different from what was claimed in
\cite{CF01}.

Before turning to the next case, let us note that the self-similar solutions
given by Eq.(\ref{3.11a})   correspond  to the choice,
\bq
\lb{A.16}
\beta(v) = v^{1/(\chi + 1)}.
\eq

\subsection*{A.2 Exact Solutions for $\alpha'(u) \beta'(v) \not= 0$}
In this case solving Eqs.(\ref{A.1a}--\ref{A.1e}) in general is very  
difficult.  Instead, we can prove the following theorems. 

\bigskip 
{\bf Theorem 1}: For any given solution, $M(u, v),\; U(u,v)$, and
$V(u, v)$ of the Einstein {vacuum} field equations, $R^{(4)}_{\mu\nu} =
0$, for the { four-dimensional spacetimes} described by the metric, 
\bq
\lb{A.17}
ds^{2}_{(4)} = - 2e^{-M}dudv + e^{-U} \left(e^{V}dx^{2} +
e^{-V}dy^{2}\right),  
\eq 
the solutions,  
\bqn
\lb{A.18}
\sigma &=& - \frac{1}{2}\left[(2 - 3\gamma^{2})U - (1 -
2\gamma^{2})\ln\left|U_{,u}U_{,v}\right| + 2\gamma^{2}M\right] +
\sigma_{0},\nb\\
r &=& e^{-U},\;\;\;\;\;\;  \phi = \frac{\gamma}{\sqrt{8\pi G}} V, 
\eqn
satisfy the Einstein-scalar field equations (\ref{A.1a}--\ref{A.1e}), where
$\gamma$ and $\sigma^{1}_{0}$ are two arbitrary constants.

The proof of the above theorem is straightforward. In fact, substituting
Eq.(\ref{A.18}) into Eqs.(\ref{A.1a}--\ref{A.1e}), we find that these
equations, in terms of $M, \; U$ and $V$,  take the
form 
\bqn
\lb{A.19a}
2U_{,uu} - U_{,u}^{2} + 2U_{,u}M_{,u} &=& V^{2}_{,u},\\
\lb{A.19b}
2U_{,vv} - U_{,v}^{2} + 2U_{,v}M_{,v} &=& V^{2}_{,v},\\
\lb{A.19c}
2M_{,uv} + 2U_{,uv} - U_{,u}U_{,v} &=& V_{,u}V_{,v},\\
\lb{A.19d}
2V_{,uv} - U_{,u}V_{,v} - U_{,v}V_{,u} &=& 0,\\
\lb{A.19e}
U_{,uv} - U_{,u}U_{,v} &=& 0,
\eqn
which are exactly the Einstein vacuum field equations for the
four-dimensional spacetimes described by the metric  (\ref{A.17}) \cite{TW90}.
Introducing a new timelike coordinate $t$ via the relation
\bq
\lb{A.20}
t = \alpha(u) - \beta(v),
\eq
we find that $t$ and $r$ are linearly independent, because now we have the
condition $\alpha'(u) \beta'(v) \not= 0$. Then, in terms of  
$t$ and $r$, the metric (\ref{A.17}) can be written in the form   
\bq
\lb{A.17a}
ds^{2}_{(4)} = - \frac{e^{\Omega(t, r)}}{\sqrt{r}}\left(dt^{2} -
dr^{2}\right) + r \left[e^{V(t, r)}dx^{2} + e^{-V(t, r)}dy^{2}\right], 
\eq 
where
\bq
\lb{A.17aa}
\Omega \equiv - M - \frac{1}{2}U - \ln\left|2\alpha'(u)\beta'(v)\right|. 
\eq
Hence, in terms of $\Omega$   the Einstein vacuum field
equations (\ref{A.19a}--\ref{A.19e}) take the form 
\bqn
\lb{A.19aa}
\Omega_{,r} &=& \frac{r}{2}\left(V^{2}_{,r} + V^{2}_{,t}\right),\\ 
\lb{A.19ab}
\Omega_{,t} &=& {r} V_{,r}V_{,t},\\
\lb{A.19ac}
\Omega_{,rr} &-& \Omega_{,tt} = \frac{1}{2} \left(V^{2}_{,t} -
V^{2}_{,r}\right),\\
\lb{A.19ad}
2V_{,rr} &+& \frac{1}{r}V_{,r} - V_{,tt} = 0.
\eqn
On the other hand, substituting Eq.(\ref{A.17aa}) into Eq.(\ref{A.18}), we find
that   
 \bqn
\lb{A.18a}
\sigma &=& \gamma^{2} \Omega +
\frac{1}{2}\ln\left|\alpha'(u)\beta'(v)\right|  + \sigma _{0},\nb\\
 t &=& \alpha(u) - \beta(v),\;\;\;\ r = \alpha(u) + \beta(v),\nb\\
\phi &=& \frac{\gamma}{\sqrt{8\pi G}} V, \;\;\; (\alpha'\beta' \not= 0).
\eqn
By properly imposing physical conditions \cite{PSW96}, the metric
given by Eq.(\ref{A.17}) can be interpreted as representing cylindrically
symmetric spacetimes with one of the two coordinates $x$ and $y$ being  the
angular coordinate and  the other being the axial coordinate. But, if we take
the values of $x,\; y$   in the range $ - \infty < x,\;  y < \infty$, the
metric is usually  considered  as representing cosmological models
\cite{Kramer80}. However, since here we are mainly interested in obtaining 
exact solutions of the ($2+1$)-dimensional metric (\ref{2.6}), we shall not be
concerned with the physical interpretation of the metric (\ref{A.17}).

It is interesting to note that if we make the replacement
\bq
\lb{A.18aa}
t = i z, \;\;\;\; 
\eq
in Eqs.(\ref{A.19aa}--\ref{A.19ad}) we find that
\bqn
\lb{A.19ba}
\Omega_{,r} &=& \frac{r}{2}\left(V^{2}_{,r} - V^{2}_{,z}\right),\\ 
\lb{A.19bb}
\Omega_{,z} &=& {r} V_{,r}V_{,z},\\
\lb{A.19bc}
\Omega_{,rr} &+& \Omega_{,zz} = -\frac{1}{2} \left(V^{2}_{,z} +
V^{2}_{,r}\right),\\
\lb{A.19bd}
2V_{,rr} &+& \frac{1}{r}V_{,r} + V_{,zz} = 0,
\eqn
which are exactly the Einstein vacuum equations for four-dimensional
spacetimes with axisymmetry \cite{Kramer80} 
\bq
\lb{A.22a}
d\tilde{s}^{2}_{(4)} = e^{\Omega - V}\left(dr^{2} + dz^{2}\right) +
r^{2}e^{-V}d\psi^{2} - e^{V}dt^{2}.
\eq

{\bf Theorem 2} \cite{TW91}: For any given solutions
$M(u,v), \; V(u,v)$, and $ U(u,v)$ of the Einstein vacuum field equations
(\ref{A.19a}--\ref{A.19e}), those with 
 \bqn 
\lb{A.21}
 F(u,v) &=& (1 - \delta^{2})\left(\frac{3}{2}U -
\ln\left|U_{,u}U_{,v}\right|\right) + \delta^{2}M + F_{0},\nb\\ 
\Phi(u, v) &=& \frac{\delta}{\sqrt{16\pi G}}V, 
\eqn
where $\delta$ and $F_{0}$ are two arbitrary constants,
satisfy the Einstein-scalar field equations, $\bar{R}_{\mu\nu}^{(4)} =
8\pi G \Phi_{,\mu} \Phi_{,\nu}$, of the {\em four-dimensional spacetimes with
 plannar symmetry},   
\bq
\lb{A.22}
{d\bar{s}}^{2}_{(4)} = - 2e^{-F(u,v)}dudv + e^{-U(u, v)}\left(dx^{2} +
dy^{2}\right). 
\eq

The combination of Eqs.(\ref{A.18}) and (\ref{A.21}) yields,
\bqn
\lb{A.23}
\sigma(u,v) &=& \frac{1}{2\delta^{2}}\left\{
(3\gamma^{2} - 2\delta^{2})U + (\delta^{2} -
2\gamma^{2})\ln\left|U_{,u}U_{,v}\right|\right\} -
\frac{\gamma^{2}}{\delta^{2}}F + \sigma^{1}_{0},\nb\\
r(u, v) &=& e^{-U},\;\;\;\;\;\;\;\;\;
\phi(u,v) = \frac{\sqrt{2} \gamma}{\delta}\Phi.
\eqn
Since the vacuum spacetimes of Eqs.(\ref{A.17}), (\ref{A.22a}) and the massless
scalar spacetimes of Eq.(\ref{A.22}) have been extensively studied,
and many analytical solutions are known \cite{TW91,Kramer80,Verd85,Que89}, 
we can use the above two theorems to construct solutions of the
Einstein-scalar field equations in ($2+1$)-dimensional spacetimes described by
the metric  (\ref{2.6}).   In particular, Eq.(\ref{A.19d}) has the
general solutions, 
\bqn
\lb{A.24}
V &=& c t + d \ln(r) + \frac{1}{2}\sum^{N}_{k =
1}{g_{k}\ln\left(\sigma_{k}\right)}
+ \sum^{N}_{k =
1}{h_{k}\mbox{Arcos}\left[\frac{2t_{k}\sigma^{1/2}_{k}}{r(1
+ \sigma_{k})}\right]}\nb\\
& & + \int^{\infty}_{0}{\left[A(\omega)\sin(\omega t) + B(\omega)\cos(\omega
t)\right]J_{0}\left(\omega r\right)} d\omega\nb\\
& & + \int^{\infty}_{0}{\left[C(\omega)\sin(\omega t) +
D(\omega)\cos(\omega t)\right]N_{0}\left(\omega r\right)}d\omega,    
\eqn
where $c,\; d,\; g_{k},\; h_{k},\; A(\omega),\; B(\omega),\;
C(\omega),\; D(\omega)$ and $ \omega$ are arbitrary real constants, and $N$ is
an integer. The functions  $J_{0}\left(\omega r\right)$ and $N_{0}\left(\omega
r\right)$ denote the Bessel and Neumann functions of  zero order, respectively,
while $\sigma_{k}$ can be chosen as $\sigma^{+}_{k}$ or $
\sigma^{-}_{k}$, which are defined by \cite{Verd85},
\bqn
\lb{A.25}
\sigma^{\pm}_{k} &\equiv& L_{k} \pm \left(L^{2}_{k} - 1\right)^{1/2},\nb\\
L_{k} &=& \frac{t^{2}_{k} + w_{k}^{2}}{r^{2}} + \left[1 -
\frac{2\left(t^{2}_{k} - w_{k}^{2}\right)}{r^{2}} + \frac{\left(t^{2}_{k} +
w_{k}^{2}\right)^{2}}{r^{4}}\right]^{1/2},
\eqn
where 
\bq
\lb{A.26}
t_{k} \equiv t^{0}_{k} - t,\;\; ( k = 1, 2, ..., N),
\eq
with $t^{0}_{k}$ and $w_{k}$ being real constants. From the above
expressions it can be shown that
\bq
\lb{A.26aa}
0 < \sigma^{-}_{k} < 1,\;\;\;\;\;
1 < \sigma^{+}_{k} < \infty.
\eq
The third and fourth terms in Eq.(\ref{A.24})
are usually called generalized soliton solutions.  

The corresponding general solutions for $\Omega$ are not known, but, in some
particular cases one can find it from Eqs.(\ref{A.19ba}) and (\ref{A.19bb}) by
quadratures. For example, when  $d$ and $g_{k}$'s are the only constants that
are different from zero, it is given by,
\bq
\lb{A.27}
e^{\Omega} = f_{0}r^{(d^{2} - g^{2})/2}\prod^{N}_{k =
1}{\frac{\sigma_{k}^{g_{k}(2g_{k} + d - g)/2}}{\left[(1 -
\sigma_{k})^{2}H_{k}\right]^{g^{2}_{k}/4}}}
\prod^{N}_{i,j = 1 (i > j)}{H_{ij}^{g_{i}g_{j}/2}},
\eq
where $f_{0}$ is a constant, and
\bqn
\lb{A.28}
g &\equiv& \sum^{N}_{k=1}{g_{k}},\;\;\;\;
H_{k} \equiv (1 - \sigma_{k})^{2} + \frac{16w^{2}_{k}\sigma^{2}_{k}}{r^{2}(1 -
\sigma_{k})^{2}},\nb\\
H_{ij} &\equiv& \left[(\sigma_{i} + \sigma_{j})r^{2} 
- \frac{8 t_{i}t_{j}\sigma_{i}\sigma_{j}}{(1+\sigma_{i})(1+
\sigma_{j})}\right]^{2}
- \left[\frac{8 w_{i}w_{j}\sigma_{i}\sigma_{j}}{(1+\sigma_{i})(1+
\sigma_{j})}\right]^{2}.
\eqn

On the other hand, if $d$ and $h_{k}$'s are the only non-zero constants in
Eq.(\ref{A.24}), the corresponding $\Omega$ is given by
\bqn
\lb{A.27a}
e^{\Omega} &=& f_{0}r^{(d^{2} + 2h)/2}\prod^{N}_{i,j = 1 (i >
j)}{\left(\frac{A^{-}_{ij}}{A^{+}_{ij}}\right)^{h_{i}h_{j}/2}}
\times \exp\left\{d\sum^{N}_{k =
1}{h_{k}\mbox{Arcos}\left[\frac{2t_{k}\sigma^{1/2}_{k}}{r(1
+ \sigma_{k})}\right]}\right\}\nb\\
& & \times \prod^{N}_{k =1}
{\frac{\left(A_{k}\sigma^{2}_{k}\right)^{h_{k}(h_{k} -
1)/4}}{H_{k}^{h_{k}(h_{k} - 2)/4}\left[r^{2}(1 -
\sigma_{k})\right]^{h^{2}_{k}/2}}}, \eqn
where $H_{k}$ is defined by Eq.(\ref{A.28}), and
\bqn
\lb{A.28a}
h& \equiv& \sum^{N}_{k = 1}{h_{k}},\;\;\;\;\;\;\;\;\;\;
A_{k} \equiv (t^{2}_{k} -
w^{2}_{k} - r^{2})^{2} + 4 w^{2}_{k}t^{2}_{k},\nb\\
A^{\pm}_{ij} &\equiv& (1 - \sigma^{2}_{i})(1 -
\sigma^{2}_{j})\left\{r^{2}(\sigma_{i} + \sigma_{j})\right.\nb\\
& & \left. -
8\sigma_{i}\sigma_{j}\left[\frac{t_{i}t_{j}}{(1 + \sigma_{i})(1 +
\sigma_{j})} \pm \frac{w_{i}w_{j}}{(1 - \sigma_{i})(1 -
\sigma_{j})}\right]\right\}. 
\eqn

Moreover, the separation of variables of Eq.(\ref{A.19bd}) yields the
solutions \cite{Que89},
\bq
\lb{A.29}
V_{Axial}(r, z) = \sum^{\infty}_{n =
0}{a_{n}\frac{P_{n}(\cos\theta)}{\rho^{n+1}}},
\eq
where $a_{n}$ are constants, and
\bq
\lb{A.30}
\rho \equiv r^{2} + z^{2},\;\;\;\;\;
\cos\theta = \frac{z}{\rho}.
\eq
Then, the integration of Eqs.(\ref{A.19ba}) and (\ref{A.19bb}) gives,
\bq
\lb{A.31}
\Omega_{Axial}(r, z) = \sum^{\infty}_{l,n =
0}\frac{a_{n}a_{l}B_{ln}}{\rho^{l+n+2}}\left(P_{l+1} P_{n+1} 
- P_{l} P_{n}\right), 
\eq
where  $P_{n}$ represents the Legendre polynomial
of order $n$, and
\bq
\lb{A.32}
 B_{ln} \equiv \frac{(l+1)(n+1)}{2(l+n+2)}.
\eq

\end{document}